\documentclass[11pt]{article}

\usepackage[a4paper,margin=2.8cm,bottom=3.8cm]{geometry}

\usepackage{graphicx}
\usepackage{amssymb}
\usepackage{amsmath}
\usepackage{amsthm}
\usepackage{mathrsfs}
\usepackage{url}
\usepackage{color}
\usepackage{longtable}
\usepackage{rotating}
\usepackage{authblk}
\usepackage{multirow}
\usepackage[authoryear]{natbib}

\usepackage{array}
\newcolumntype{P}[1]{>{\centering\arraybackslash}p{#1}}
\usepackage{bbm}

\usepackage{float}
\usepackage{caption}
\usepackage[titletoc,title]{appendix}

\begin{document}

\title{Multi-Level Order-Flow Imbalance in a Limit Order Book}

\author[1]{Ke Xu\thanks{Corresponding author. Email: \texttt{xuke\_e@hotmail.com.}}}
\author[1]{Martin D. Gould}
\author[1]{Sam D. Howison}
\affil[1]{Mathematical Institute, University of Oxford, Oxford OX2 6GG, UK}

\maketitle

\begin{abstract}We study the \emph{multi-level order-flow imbalance (MLOFI)}, which is a vector quantity that measures the net flow of buy and sell orders at different price levels in a limit order book (LOB). Using a recent, high-quality data set for 6 liquid stocks on Nasdaq, we fit a simple, linear relationship between MLOFI and the contemporaneous change in mid-price. For all 6 stocks that we study, we find that the out-of-sample goodness-of-fit of the relationship improves with each additional price level that we include in the MLOFI vector. Our results underline how order-flow activity deep into the LOB can influence the price-formation process.\end{abstract}


\section{Introduction}\label{sec:introduction}

In most modern financial markets, trade occurs via a continuous double-auction mechanism called a \emph{limit order book (LOB)}. In an LOB, price changes are a consequence of the actions and interactions of many heterogeneous traders, each of whom submits and/or cancels orders. Throughout the past decade, the question of how price changes emerge from the complex interplay of order flows has attracted considerable attention from academics (see \cite{Gould:2013limit} for a survey and \citet{Bouchaud:2018trades} for a textbook treatment), because a deeper understanding of the origins and nature of price changes provides a conceptual bridge between the microeconomic mechanics of order matching and macroeconomic concept of price formation. The same topic is also important in many practical situations, including market making \citep{sandaas2001adverse}, designing optimal execution strategies \citep{alfonsi2010optimal}, and minimizing market impact \citep{donier2015fully}.

During the past 20 years, a wide variety of research efforts have formulated and tested models regarding specific aspects of price formation in an LOB. We review a selection of such studies in Section \ref{sec:price_formation}. Although these models provide important insights into the delicate interplay between order flow, liquidity and price formation, they typically suffer from an important drawback: They are extremely complex. To address this problem, some authors have sought to uncover and fit simple statistical relationships that explain price formation in an LOB in terms of easily understood and easily measurable inputs, such as net order flow. Our work contributes to this emerging thread of the literature.

In a seminal work on this topic, \citet{Cont:2014price} proposed a scalar quantity called the \emph{order flow imbalance (OFI)}, which measures the net order flow at the best quotes during a specified time window (we provide a detailed description of OFI in Section \ref{sec:ofi}). Using trades-and-quotes (TAQ) data on 50 stocks from the S\&P 500, the authors performed an ordinary least squares (OLS) regression to fit a simple linear relationship between the OFI and the contemporaneous change in mid-price. These regressions achieve a mean $R^2$ statistic of about $65\%$, and are statistically significant at the $95\%$ level in about $98\%$ of the samples. Cont \emph{et al.} thereby concluded that this simple linear relationship provides a powerful link between order flow and price formation.

In Appendix B3 of their paper, \citet{Cont:2014price} briefly extended their discussion by including not only the OFI at the best quotes, but also a similar measure of net order flow at the first 5 price levels on each side of the LOB. The authors performed a multivariate OLS regression to fit a linear relationship between these 5 net order flows and the contemporaneous change in mid-price. Including these additional price levels only slightly improved the goodness-of-fit beyond that achieved by their OFI regressions, which led the authors to conclude that order-flow activity beyond the best quotes has very little influence on price changes.

In this paper, we build on the work of \citet{Cont:2014price} by performing a more detailed analysis of the relationship between the net order flow at the first $M$ price levels on each side of an LOB and the contemporaneous change in mid-price. We call this $M$-dimensional vector the \emph{multi-level order-flow imbalance (MLOFI)}. We perform empirical calculations using a recent, high-quality, high-frequency data set that describes all LOB activity during 2016 for 6 liquid stocks on Nasdaq. We first consider the case $M=1$, and use OLS regression to fit a linear relationship between the OFI and the contemporaneous change in mid-price. Consistently with Cont \emph{et al.}, we find the fitted relationship to be strongly statistically significant. We then extend our analysis to include net order flow at price levels deeper into the LOB. When we use OLS regression to fit the MLOFI relationship (as did Cont \emph{et al.}), we find that the statistical significance of the individual parameters is relatively weak. However, for all 6 stocks in our sample, we also uncover strong sample correlations between the net order flow at different price levels. We therefore argue that OLS regression is likely to produce unstable fits of the MLOFI relaltionship.

To address this problem, we use Ridge regression to perform similar empirical fits. When using Ridge regression, we reach a very different conclusion: We find that even the parameters that correspond to price levels deep into the LOB are strongly statistically significant. To assess the goodness-of-fit of our regressions, we study both the $R^2$ statistic and the root mean squared error (RMSE) of the fitted relationships. By comparing the in-sample and out-of-sample performance, we uncover evidence to suggest that the OLS regressions over-fit the data for some stocks. This provides further evidence to suggest that multivariate OLS is not well-suited for performing the regressions with the MLOFI vectors.

By analyzing the out-of-sample RMSE of the Ridge regression fits, we find that including net order flow deeper into the LOB improves the goodness-of-fit of the MLOFI regressions for all of the stocks in our sample, with an improvement of about $65$--$75\%$ for large-tick stocks and about $15$--$30\%$ for small-tick stocks. We argue that in many practical applications, improvements of this magnitude are economically meaningful.

The paper proceeds as follows. In Section \ref{sec:price_formation}, we discuss the core concepts related to price formation in an LOB and review a selection of studies relevant to our work. In Section \ref{sec:mlofi}, we introduce multi-level order-flow imbalance (MLOFI). In Section \ref{sec:data}, we describe the data that we use for our empirical calculations. We present our main empirical result in Section \ref{sec:results} and discuss our findings in Section \ref{sec:discussion}. Section \ref{sec:conclusions} concludes.

\section{Price Formation in an LOB}\label{sec:price_formation}

\subsection{Order Flow, Liquidity and Price Formation}\label{sec:order_flow}

An LOB can be regarded as a system of queues at different prices. The prices at which the queues reside are called \textit{price levels}. The \emph{level-1 bid-price} (often simply called the \emph{bid price}) refers to the highest price among buy limit orders. The \emph{level-2 bid-price} refers to the second-highest price among buy limit orders, and so on. The \emph{level-1 ask-price} (often simply called the \emph{ask price}) refers to the lowest price among sell limit orders. The \emph{level-2 ask-price} refers to second-lowest price among sell limit orders, and so on.

For $m\geq1$, and at a given time $t$, let $b^m(t)$ denote the level-$m$ bid-price and let $a^m(t)$ denote the level-$m$ ask-price.\footnote{In many other studies of LOBs, it is customary to write $b(t)$ to denote the bid-price and $a(t)$ to denote the ask price. In our notation, we write $b^1(t)$ and $a^1(t)$.} Let $s(t): = a^1(t) - b^1(t)$ denote the \emph{bid--ask spread}, and let $P(t): = (a^1(t) + b^1(t))/2$ denote the \emph{mid-price at time $t$}. Throughout this paper, we measure all such quantities when including the effect of any order-flow activities at time $t$. For example, if a new buy limit order arrives at time $t$ with a price $p$ that is strictly greater than the previous level-1 bid-price, then we regard the level-1 bid-price at time $t$ to be equal to the price of this newly arriving limit order, $b^1(t)=p$.

For a given $\Delta t >0$, let $\Delta P(t,t+\Delta t) := P(t+\Delta t) - P(t)$ denote the change in mid-price that occurs between times $t$ and $t+\Delta t$. The core aim of this paper is to investigate the relationship between the net order flow at a given set of prices levels and the contemporaneous change in mid-price.

\subsection{Literature Review}

In recent years, the increasing availability of high-quality, high-frequency data sets from LOBs has stimulated a wide range of empirical and theoretical studies that seek to address this relationship. The literature on this topic is vast, and spans many different disciplines, including economics, physics, mathematics, statistics and psychology. For detailed surveys of work in this field, see \citet{Bouchaud:2009digest} and \cite{Gould:2013limit}. In the remainder of this section, we review a selection of publications most relevant to our own work.

\subsubsection{Models of LOB State}

Early models of price formation in an LOB typically assumed that order flows were governed by simple, independent stochastic processes. By tracking how such order flows impact the state of an LOB over time, such models can be used to generate predictions of $\Delta P(t,t+\Delta t)$. \citet{Smith:2003statistical} introduced a model in which limit order arrivals, market order arrivals and cancellations all occur as mutually independent Poisson processes with fixed rate parameters. \citet{Cont:2010stochastic} extended this model by allowing the rates of limit order arrivals and cancellations to vary across prices. \citet{Huang:2015simulating} studied a model in which order arrival rates depend on the state of the LOB. \citet{Mike:2008empirical} considered an LOB model in which order flows exhibit long-range autocorrelations, in agreement with empirical data (see Section \ref{sec:long_memory}).

Although these models produce many good predictions of the long-run statistical properties of LOB state (see, e.g., \citet{Farmer:2005predictive}), they typically produce poor predictions of $\Delta P(t,t+\Delta t)$. We suggest two possible reasons for why this is the case. First, they generate price series that are extremely complex and very sensitive to the underlying modelling assumptions. This makes it difficult to relate their outputs to those of real-world LOBs, which, in turn, makes it difficult to understand how to improve them. Second, although these models use order flows as their inputs, they do not seek to reflect the strategies implemented by traders. Therefore, they fail to capture many important correlations and feedback loops in their inputs, which causes them to produce unrealistic dynamics of $\Delta P(t,t+\Delta t)$.

\subsubsection{Models of the Long Memory of Order Flow}\label{sec:long_memory}

Another thread of literature on this topic arises from an important empirical observation that was made independently by \citet{Lillo:2004long} and \citet{Bouchaud:2004fluctuations}: Time series of market order arrival signs (i.e., whether each arriving market order is to buy or to sell) exhibit long memory. Given that the mid-price $P(t)$ can be regarded as a superposition of the impact of all previous market order arrivals, this raised the important question of why price changes were not highly predictable (see, e.g., \citet{Farmer:2006market} and \citet{Bouchaud:2009digest} for wider discussion).

To help resolve this conundrum, \citet{Lillo:2004long} introduced a model in which price impact is regarded as variable and permanent. In this approach, price efficiency is maintained by a corresponding liquidity imbalance that co-moves with the imbalance of buy versus sell market orders. \citet{Bouchaud:2004fluctuations} proposed an alternative solution to the puzzle, by introducing a model in which price impact is fixed and temporary, and governed by a so-called ``propagator'' function.

\citet{Taranto:2018linearI} introduced an extended propagator model that partitions market orders according to whether or not they cause an instantaneous change in the mid-price. By modelling order flow as an exogenous process, the authors formulated empirical predictions about mid-price changes for the 50 most traded stocks on the NYSE and Nasdaq. They reported that their model's predictions far outperform those of standard propagator models, especially for large-tick stocks. \citet{Taranto:2018linearII} further extended this approach by modelling order-flow events via a mixture transition distribution \citep{Raftery:1985model}. They reported that this approach produces better predictions for changes in the mid-price, which they argued was due to its ability to capture the conditional correlations between the different types of order flow. The authors also performed an out-of-sample analysis of their model's goodness-of-fit to verify that its improvements were not a consequence of over-fitting.

\subsubsection{Models of Order Queues at the Bid- and Ask-Prices}

\citet{Cont:2013price} introduced a model in which the numbers of orders at the level-1 bid- and ask-prices are governed by independent diffusion processes. By studying this model in the hydrodynamic limit, in which stochastic fluctuations are dominated by deterministic flows, the authors obtained analytical expressions for several quantities of interest, such as the distribution of times between price changes and the distribution of changes in the mid-price.

% \citet{Gareche:2013fokker} presented empirical results to illustrate that the number of orders at the level-1 bid- and ask-prices are strongly coupled.

\citet{Avellaneda:2011forecasting} extended the model from \citet{Cont:2013price} by introducing an explicit correlation between the number of orders at the level-1 bid- and ask-prices. The authors solved their model to deduce a simple, closed-form predictor for the direction of the next mid-price movement, and reported that their model produces reasonably good predictions of mid-price movements in real LOBs.

\subsubsection{Trade Imbalance and Order-Flow Imbalance}\label{sec:ofi}

Although all of the above models provide interesting insights into the delicate interplay between order flow, liquidity, and price formation, they all suffer a common drawback: They are extremely complex. Motivated by this issue, several authors have proposed simple statistical models that seek to explain price formation in an LOB in terms of very simple inputs, such as net order flow. To date, such models have typically considered one of two input variables: trade imbalance or order-flow imbalance (OFI).

On a given trading day, fix a uniformly spaced time grid
\[ t_0 < t_1 < \ldots < t_{k-1} < t_k < \ldots < t_K.\]
For a given time interval $\left(t_{k-1}, t_{k}\right]$, let $M(t_{k-1},t_k)$ denote the total volume of buy market orders that arrive at times $t$ such that $t_{k-1} < t \leq t_{k}$. Similarly, let $N(t_{k-1},t_k)$ denote the total volume of sell market orders that arrive at times $t$ such that $t_{k-1} < t \leq t_{k}$. The \emph{trade imbalance} is given by
\begin{equation} 
TI(t_{k-1},t_k) := M(t_{k-1},t_k) - N(t_{k-1},t_k).
\end{equation}
Several empirical studies have sought to uncover and fit simple statistical relationships between the trade imbalance $TI(t_{k-1},t_k)$ and the contemporaneous change in mid-price $\Delta P(t_{k-1}, t_{k})$. By aggregating data from the 116 most frequently traded US stocks on the NYSE during 1994 to 1995, \citet{plerou2002quantifying} reported that the average mid-price change $\langle\Delta P(t_{k-1}, t_{k})\rangle$ approximately follows a hyperbolic tangent function of trade imbalance. \citet{potters2003more} studied exchange-traded funds that tracked Nasdaq and the S\&P500 during 2002, and reported a similar relationship between trade size and the corresponding price change. \citet{gabaix2006institutional} studied the 100 largest stocks on the NYSE from 1994 to 1995, and reported that the mean mid-price logarithmic return over a 15-minute interval scales approximately as the square-root of the trade imbalance. \citet{bouchaud2010price} also reported a similar (concave) power-law dependence between trade imbalance and price return, but reported that the power-law exponent increases with the length of the time interval over which returns are measured.

\citet{eisler2012price} performed an empirical study of price impact, based on market order arrivals, limit order arrivals and limit order cancellations, for 14 liquid stocks traded on Nasdaq between 3 March and 19 May 2008. By analyzing the cross correlation between order flows and the corresponding price movements, the authors concluded that, on average, limit orders have similar price impact to market orders. Importantly, this work suggests that studying all order flows, including limit order arrivals and cancellations, could provide a fuller statistical picture of price impact in an LOB than the one provided by considering only market order arrivals.

% Importantly, this insight suggests that the approach of looking solely at trades, and ignoring limit order arrivals, misses an important part of the story.

\cite{Cont:2014price} made a similar observation when investigating relationships between trade imbalance and the contemporaneous change in mid-price: $TI(t_{k-1},t_k)$ considers only market orders, and omits the possible influence of limit order arrivals and cancellations on the mid-price. To help address this problem, the authors proposed a new quantity, which they called the \textit{order-flow imbalance (OFI).} OFI quantifies the net order-flow imbalance at the best quotes, in terms of the market order arrivals, limit order arrivals and cancellations.

Let $\tau_n$ denote the time of the $n^{\text{th}}$ order arrival or cancellation. In this way, $b^m(\tau_n)$ denotes the level-$m$ bid-price and $a^m(\tau_n)$ denotes the level-$m$ ask-price at the time $\tau_n$. Recall from Section \ref{sec:order_flow} that when we measure $b^m(t)$ and $a^m(t)$, we always include the effect of any order-flow activities at time $t$. Therefore, we measure all values of $b^m(\tau_n)$ and $a^m(\tau_n)$ immediately \emph{after} applying the effect of the $n^{\text{th}}$ order arrival or cancellation.

Let $q^m(\tau_n)$ denote the total size of all orders at the level-$m$ ask-price, and let $r^m(\tau_n)$ denote the total size of all orders at the level-$m$ bid-price, again measured \emph{after} applying the effect of the $n^{\text{th}}$ order arrival or cancellation. Between any two consecutive order events (for concreteness, we consider the times $\tau_{n-1}$ and $\tau_n$), let
\begin{equation}
e_n := \Delta {W(\tau_n)} - \Delta {V(\tau_n)},
\end{equation}
where
\begin{equation}\label{eq:Vb}
\Delta {W(\tau_n)} =\begin{cases}
    r^1(\tau_n), & \text{if } b^1(\tau_n) > b^1(\tau_{n-1}),\\
    r^1(\tau_n) - r^1(\tau_{n-1}), & \text{if } b^1(\tau_n) = b^1(\tau_{n-1}),\\
    - r^1(\tau_{n-1}), & \text{if } b^1(\tau_n) < b^1(\tau_{n-1});\\
  \end{cases}
\end{equation}
and
\begin{equation}\label{eq:Va}
\Delta {V(\tau_n)} =\begin{cases}
    - q^1(\tau_{n-1}), & \text{if } a^1(\tau_n) > a^1(\tau_{n-1}),\\
    q^1(\tau_n) - q^1(\tau_{n-1}), & \text{if } a^1(\tau_n) = a^1(\tau_{n-1}),\\
    q^1(\tau_n), & \text{if } a^1(\tau_n) < a^1(\tau_{n-1}).\\
  \end{cases}
\end{equation}

Using this notation, the OFI for a given time interval $\left( t_{k-1}, t_k\right]$ is given by the sum of the $e_n$ from all order arrivals and cancellations that occurred during the time interval:

\begin{equation}\label{eq:ofi_e_n}
OFI(t_{k-1}, t_{k}) = \sum_ {\{n \vert t_{k-1} < \tau_n \leq t_k \}} {e_n}.
\end{equation}

% Observe that equations \eqref{eq:Vb} and \eqref{eq:Va} partition the space of all possible order-flow activity. Assume that the order-flow activity at time $\tau_n$ occurred on the buy-side of the LOB. First consider $W(\tau_n)$. This order-flow activity must have caused one of the following three scenarios:
%
%\begin{itemize}
%\item The level-1 bid-price increased (which occurs when the order-flow activity is a limit order arriving inside the bid--ask spread);
%\item The level-1 bid-price did not change;
%\item The level-1 bid-price decreased (which occurs when the order-flow activity removed all limit orders at the previous level-1 bid-price).
%\end{itemize}
%
% Next, consider $V(\tau_n)$. The order-flow activity at time $\tau_n$ occurred on the buy-side of the LOB, so, by definition, $a^1(\tau_n) = a^1(\tau_{n-1})$ and $q^1(\tau_n) = q^1(\tau_{n-1})$. Therefore, $V(\tau_n)=0$. Symmetric arguments hold for activity on the sell-side of the LOB.

% This depth-change based construction of OFI is a close approximation to the order-size based construction defined in \eqref{sec:ofi}. \citep{Cont:2014price} used this defintation to construct OFI in their study due to the lack of information on every single order event in TAQ database. In this study, we use the same method (depth-based construction)  to implement our proposed indicator MLOFI because only depth (instead of order) information in an LOB is available to all the market participants in a real trading environment.

The simple, scalar-valued quantity $OFI(t_{k-1}, t_{k})$ measures the direction and magnitude of the net order flow at the bid- and ask-prices during the time interval $\left(t_{k-1}, t_k\right]$. The OFI is positive if and only if the total size of buy limit order arrivals at the bid-price, sell limit order cancellations at the ask price, and buy market order arrivals is greater than the total size of sell limit order arrivals at the ask-price, buy limit order cancellations at the bid-price, and sell market order arrivals. In other words, $OFI(t_{k-1}, t_{k})$ is positive if and only if the aggregated buying pressure at the best quotes is greater than the aggregated selling pressure at the best quotes. As \cite{Cont:2014price} argued, the mechanics of LOB trading suggest that the larger the value of $OFI(t_{k-1}, t_{k})$, the greater the probability that $\Delta P(t_{k-1}, t_{k})$ will be positive. Conversely, the more negative the value of $OFI(t_{k-1}, t_{k})$, the greater the probability that $\Delta P(t_{k-1}, t_{k})$ will be negative.

To assess the relationship between $OFI(t_{k-1}, t_{k})$ and $\Delta P(t_{k-1}, t_{k})$ in more detail, \citet{Cont:2014price} chose 50 stocks at random from the S\&P 500 index. For each of the chosen stocks, they used the TAQ data set for the month of April 2010 to estimate\footnote{Because the TAQ data set does not contain order-flow information, the authors estimated the values of $OFI(t_{k-1}, t_{k})$ by observing the changes in volume at the bid- and ask-prices at regularly spaced points in (calendar) time. By contrast, the data that we study describes all order-flow activity, and thereby enables us to calculate net order flow exactly.} the values of $OFI(t_{k-1}, t_{k})$, and compared them with the contemporaneous values $\Delta P(t_{k-1}, t_{k})$.

For each stock and each trading day, the authors first divided the trading day into a sequence of $I$ consecutive time windows with equal length. Let $T_i$ denote the start time of the $i^{\text{th}}$ window, such that a whole trading day may be partitioned as
\begin{equation}\label{eq:T_grid}T_0 < T_1 < \ldots < T_{i-1} < T_i < \ldots < T_I.\end{equation}
For their empirical calculations, the authors set
\[\Delta T := T_i - T_{i-1} = 30\text{ minutes,}\]
which divides their period of study on each trading day (i.e. 10:00 to 15:30) into $I=11$ uniformly spaced, non-overlapping windows. Given this uniform grid, the authors sub-divided each time window into $K$ uniformly spaced sub-windows by applying a second time grid
\[t_{i,0} < t_{i,1} < \ldots < t_{i,k-1} < t_{i,k} < \ldots < t_{i,K},\]
such that $t_{i,0} = T_i$ and $t_{i,K} = T_{i+1} = t_{i+1,0}$. For their main empirical calculations, the authors set
\[\Delta t := t_{i,k} - t_{i,k-1} = 10\text{ seconds},\]
which divides each time window into $K=180$ uniformly spaced, non-overlapping time intervals. For each such time interval, the authors measured $OFI(t_{i,k-1}, t_{i,k})$ and the change in mid-price $\Delta P(t_{i,k-1}, t_{i,k})$, then used ordinary least-squares (OLS) regression to fit the linear relationship
\begin{equation}\label{eq:ofi_regression}
\Delta P(t_{i,k-1}, t_{i,k}) = \alpha + \beta OFI(t_{i,k-1}, t_{i,k}) + \varepsilon,
\end{equation}
where $\alpha$ is the intercept coefficient, $\beta$ is the slope coefficient, and $\varepsilon$ is a noise term that summarizes the influences of all the other factors not explicitly considered in the model. For their fitted regressions, they found that $\alpha$ was statistically significant in only $10\%$ of the cases they examined, but that $\beta$ was statistically significant in $98\%$ of the cases they examined. They therefore argued that a simple model of the form
\begin{equation}\label{eq:cont_linear_model}
\Delta P(t_{i,k-1}, t_{i,k}) \approx \beta OFI(t_{i,k-1}, t_{i,k})
\end{equation}
provides a good fit to the data. The authors also considered non-linear relationships with higher-order terms, but reported little improvement over this simple linear model.

% The authors further investigated the slope parameter $\beta$. By performing regression analysis between the average market depth $D$ (i.e., the time-average of $q^1(\tau_n)$ and $r^1(\tau_n)$) and the slope parameter $\beta$ for each stock, they reported the linear relationship $\beta \approx 1 / 2D$. They thereby argued that the slope parameter can be interpreted as the inverse of the mean liquidity available at the best quotes.

% By adding a higher order term of OFI, i.e. $OFI|OFI|$, into the linear regression model, the authors conclude the linear relationship between OFI and the price change, as the estimated parameter of the higher order term is statistically significant in only 17\% of samples.

The authors also performed the same OLS regressions using trade imbalance as the input variable, and used the $R^2$ statistic to compare the (in-sample) goodness-of-fit to that achieved by OFI. For their fits using trade imbalance, they found an average $R^2$ of about $32\%$; for their fits using OFI, they found an average $R^2$ of about $65\%$. They thereby concluded that OFI far outperforms trade imbalance in terms of its explanatory power for contemporaneous changes in mid-price.

\subsubsection{Modelling Price impact as a Dynamical and Latent Variable}

In a very recent working paper, \citet{Mertens:2019liquidity} extended the OFI framework from \citet{Cont:2014price} by modelling price impact as a dynamical and latent variable that is equal to the product of three components: a daily price-impact component, a deterministic intra-day pattern, and a stochastic auto-regressive component. By analyzing 5 stocks traded on Nasdaq, they reported that their approach results in a large improvement in the out-of-sample goodness-of-fit compared to that achieved by Cont \emph{et al.}'s approach. We provide a comparison of our results to those of \citet{Mertens:2019liquidity} in Section \ref{sec:discussion}.

% The average t-statistics of estimated coefficients in the two regressions are correspondingly 12.08 and 5.08. This empirical result confirms that new indicator OFI has strong explanatory power, and its performance is much better than the traditional indicator TI.

\section{Multi-Level Order-Flow Imbalance}\label{sec:mlofi}

% As we discussed in Section \ref{sec:ofi}, \citet{Cont:2014price} extended the concept of trade imbalance by including not only market order arrivals, but also limit order arrivals and cancellations at the best quotes, to define the concept of OFI. By performing separate OLS regressions of each of these input variables onto the contemporaneous change in mid-price, the authors reported a much stronger relationship between OFI and the change in mid-price than between trade imbalance and the change in mid-price.

We now turn our attention to \emph{multi-level order-flow imbalance (MLOFI)}, which forms the core of our empirical analysis.

In Appendix B3 of their paper, \citet{Cont:2014price} briefly extended their discussion by studying not only the net order flow at the best quotes, but also at price levels deeper into the LOB. When studying the first $M \geq 1$ best quotes on each side of the LOB, this produces a vector-valued quantity of dimension $M$. Because their discussion of this topic is a short appendix to the paper's main focus (i.e., OFI), the authors do not provide a detailed explanation of how they construct this vector. To avoid possible ambiguities, we first provide a detailed explanation of how we do this in our own calculations.

\subsection{Calculating MLOFI}

Recall from Section \ref{sec:ofi} that $\tau_n$ denotes the time of the $n^{\text{th}}$ order arrival or cancellation, and that $b^m(\tau_n)$ denotes the level-$m$ bid-price (i.e., the price of the $m^{\text{th}}$ populated\footnote{Observe that we only count populated price levels, so it does not necessarily follow that $a^{m+1}(\tau_n)$ is exactly one tick greater than $a^m(\tau_n)$.} best quote on the buy side of the LOB) and $a^m(\tau_n)$ denotes the level-$m$ ask-price (i.e., the price of the $m^{\text{th}}$ populated best quote on the sell side of the LOB), all measured immediately after the $n^{\text{th}}$ order arrival or cancellation, for $m = 1, 2, \ldots, M$.

Between any two consecutive order-flow events (for concreteness, we consider the events at times $\tau_{n-1}$ and $\tau_n$), let
\begin{equation}\label{eq:Vbm}
\Delta W^m(\tau_n) =\begin{cases}
    r^m(\tau_n), & \text{if }b^m(\tau_n) > b^m(\tau_{n-1}),\\
    r^m(\tau_n) - r^m(\tau_{n-1}), & \text{if }b^m(\tau_n) = b^m(\tau_{n-1}),\\
    - r^m(\tau_{n-1}), & \text{if }b^m(\tau_n) < b^m(\tau_{n-1});\\
  \end{cases}
\end{equation}
and
\begin{equation}\label{eq:Vam}
\Delta V^m(\tau_n) =\begin{cases}
    - q^m(\tau_{n-1}), & \text{if }a^m(\tau_n) > a^m(\tau_{n-1}),\\
    q^m(\tau_n) - q^m(\tau_{n-1}), & \text{if } a^m(\tau_n) = a^m(\tau_{n-1}),\\
    q^m(\tau_n), & \text{if } a^m(\tau_n) < a^m(\tau_{n-1});\\
  \end{cases}
\end{equation}
for $m=1,2,\ldots,M$.

During the time interval $[t_{k-1}, t_k)$, the $m^{th}$ entry of the MLOFI vector is given by
\begin{equation} \label{eq:mlofi_sum}
MLOFI^m (t_{k-1}, t_{k}) := \sum_{\{n \vert t_{k-1} < \tau_n \leq t_k \}} {e^m(\tau_n)},
\end{equation}
where
\begin{equation}
e^m(\tau_n) = \Delta W^m(\tau_n) - \Delta V^m(\tau_n).
\end{equation}

Observe that the definition of MLOFI is very similar to the definition of OFI in equation \eqref{eq:ofi_e_n}, albeit extended to include order flow at the best $m=1,2,\ldots,M$ occupied price levels. When $M=1$, MLOFI and OFI are identical; when $M \geq 2$, MLOFI becomes a more general measure of order-flow imbalance.

To illustrate the definition of MLOFI, consider the following example with $M=2$. Assume that only one order-flow event occurs during the time window $\left(t_{k-1}, t_k\right]$, and that this event affects the buy-side of the LOB. There are 3 possible cases to consider:

\begin{enumerate}
\item If $b^2(t_k) > b^2(t_{k-1})$, then a buy limit order must have arrived. Let $p$ denote this limit order's price and let $\omega$ denote this limit order's size. There are 2 possible sub-cases:
\begin{itemize}
\item The limit order arrives inside the bid--ask spread, such that $b^1(t_{k-1}) < p < a^1(t_{k-1}$). This limit order becomes the new level-1 queue, so $b^1(t_k) = p > b^1(t_{k-1})$ and $MLOFI^1(t_{k-1}, t_{k}) = \omega$. The previous level-1 queue becomes the new level-2 queue, so $MLOFI^{2}(t_{k-1}, t_{k}) = r^2(\tau_n) = r^1(\tau_{n-1})$. Thus, the buy limit order arrival within the bid--ask spread impacts both $MLOFI^1(t_{k-1}, t_{k})$ and $MLOFI^{2}(t_{k-1}, t_{k})$.
\item The limit order arrives between the level-1 queue (i.e., the bid-price) and the level-2 queue (such that $b^2(t_{k-1}) < p < b^1(t_{k-1})$).  In this case, the limit order arrival does not cause any change to the level-1 queue, so $MLOFI^{1}(t_{k-1}, t_{k}) = 0$. The new limit order becomes a new level-2 queue, with $b^2(\tau_n) = p > b^2(\tau_{n-1})$. Therefore, $MLOFI^{2}(t_{k-1}, t_{k}) = \omega$.
\end{itemize}

\item If $b^2(\tau_n) < b^2(\tau_{n-1})$, then there are 2 possible sub-cases:
\begin{itemize}
\item A sell market order consumed the whole previous level-1 bid-queue, or the last limit order in the previous level-1 bid-queue was cancelled. Thus, $MLOFI^{1}(t_{k-1}, t_{k}) = -r^1(\tau_{n-1})$ and $MLOFI^{2}(t_{k-1}, t_{k}) = -r^2(\tau_{n-1}) = -r^1(\tau_n)$.
\item  The last limit order in the previous level-2 bid-queue was canceled, so the previous level-3 bid-queue becomes the new level-2 bid-queue. Thus, $MLOFI^{2}(t_{k-1}, t_{k}) = -r^2(\tau_{n-1})$, while $MLOFI^{1}(t_{k-1}, t_{k}) = 0$.
\end{itemize}

%Thus, the order activities at the bid are captured by both $OFI_t = -n^b(b(t-1), t-1)$ and $MLOFI^{m=2}_t = -n^b(b_2(t-1), t-1) = -n^b(b(t), t)$. In the second scenario, the last limit order in the previous level 2 queue at $t-1$ was completely cancelled, so the previous level 3 queue at $t-1$ consequently becomes the new level 2 queue. Thus, the order activities at price level 2 are only captured by $MLOFI^{m=2}_t = -n^b(b_2(t-1), t-1)$ while $OFI_t = 0$.

\item If $b^2(\tau_n) = b^2(\tau_{n-1})$, then $MLOFI^{2}(t_{k-1}, t_{k}) = r^2(\tau_n) - r^2(\tau_{n-1})$. 
\end{enumerate}

Observe that if an order-flow event causes a change of the bid-price $b^1(\tau_n)$, then the values of $b^2(\tau_n), b^3(\tau_n), b^4(\tau_n) \ldots$ all change. Similarly, if an order-flow event causes a change in the ask-price $a^1(\tau_n)$, then the values of $a^2(\tau_n), a^3(\tau_n), a^4(\tau_n), \ldots$ all change.

% For example, when a buy limit order $x$ arrives within the bid--ask spread, it creates a new level-1 bid-queue (and therefore a new bid price $b(\tau_n) = p$), a new level-2 bid-price $b^2(\tau_n) = b^1(\tau_{n-1})$, a new level-3 bid-price $b^3(\tau_n) = b^2(\tau_{n-1})$, and so on. A similar process occurs in the opposite direction when the level-1 bid-queue is depleted to 0, either via a market order arrival or a cancellation.

% This propagation effect makes the indicator $MLOFI^{m=2}$ able to capture order activity within the first level. More generally, as this propagation effect continues passing on to all the deeper levels, the indicator $MLOFI^m$ can capture information from any order activity within level $m$, as long as the order activity causes a price change at any level $n < m$.

% When the feature variable has a positive (respectively, negative) value, the aggregated buying pressure during the event time interval $[t-n, t]$ is higher (respectively, lower) than the aggregated selling pressure. Therefore, we expect a corresponding positive (respectively, negative) value of the target variable as the mid-price increases (respectively, decreases) from event time $t-n$ to $t$. Moreover, we expect that a greater absolute value of feature variable comes with a greater absolute value of target value, because a bigger buying or selling pressure leads to a bigger price move.

\subsection{An Example}\label{sec:mlofi_example}

Set $M=3$. At time $\tau_{n-1}$, consider an LOB that contains two buy limit orders, both with size 10. The first buy limit order resides at the level-1 bid-price $b^1(\tau_{n-1})=\$1.40$, and the second buy limit order resides at the level-2 bid-price $b^2(\tau_{n-1})=\$1.39$, such that
\begin{equation}
\begin{cases}
    b^1(\tau_{n-1})=\$1.40, & r^1(\tau_{n-1})=10,\\
    b^2(\tau_{n-1})=\$1.39, & r^2(\tau_{n-1})=10.\\
\end{cases}
\end{equation}

During the time interval $\left(t_{k-1}, t_k\right]$, assume that the only order-flow activity is the arrival of a single buy limit order with price $p=\$1.41$ and size $\omega=7$. This limit order becomes the new level-1 bid-queue, such that
\begin{equation}
\begin{cases}
    b^1(\tau_n)=\$1.41, & r^1(\tau_n)=7,\\
    b^2(\tau_n)=\$1.40, & r^2(\tau_n)=10,\\
    b^3(\tau_n)=\$1.39, & r^3(\tau_n)=10,\\
\end{cases}
\end{equation}
Therefore,
\begin{equation}
\begin{cases}
MLOFI^{1}(t_{k-1}, t_{k}) = \Delta W^1(\tau_n) = r^1(\tau_n) = 7, & \text{as } b^1(\tau_n) = \$1.41 > b^1(\tau_{n-1}) = \$1.40,\\
MLOFI^{2}(t_{k-1}, t_{k}) = \Delta W^2(\tau_n) = r^2(\tau_n) = 10, & \text{as } b^2(\tau_n) = \$1.40 > b^2(\tau_{n-1}) = \$1.39,\\
MLOFI^{3}(t_{k-1}, t_{k}) = \Delta W^3(\tau_n) = r^3(\tau_n) = 10, & \text{as } b^3(\tau_{n-1}) \text{ does not exist.}\\
\end{cases}
\end{equation}
Thus, $MLOFI(t_{k-1}, t_{k}) = (7,10,10)$.

% In statistical modelling, the \textit{feature variable} is the model input, which represents possible causes for a given effect, while the target variable is the model output, which represents the consequences. In this study, we investigate the relationship between order flows at deep levels in an LOB and the corresponding price change. Therefore, the proposed multi-level OFI indicators (MLOFI) are the feature variables and the corresponding price change is the target variable. In this chapter, we describe in detail the construction method of the proposed MLOFI indicators, the linear model used to describe the relationship between feature variables and target variable, the technique used to estimate the model parameters and the model evaluation methods.

% For example, a depth increase from time $t-1$ to $t$ at price level 2 on the bid side caused by a new buy limit order $x = (p_x, \omega_x, t)$ with $p_x = b_2(t-1)$ has no impact to $OFI_t$ and $OFI(t-n, t)$. To capture these order activities at deeper price levels, we need a new indicator that is not only based on depth changes at the first price level but also based on depth changes at deeper price levels.

\section{Data}\label{sec:data}

The data that we study originates from the LOBSTER database, which provides an event-by-event description of the temporal evolution of the LOB for each stock listed on Nasdaq. The LOBSTER database contains very detailed information regarding the temporal evolution of the relevant LOBs. For a detailed introduction to LOBSTER, see \url{http://LOBSTER.wiwi.hu-berlin.de} and \citet{Bouchaud:2018trades}.

On the Nasdaq platform, each stock is traded in a separate LOB with price--time priority, with a lot size of 1 stock and a tick size of $\$0.01$. Therefore, all orders must arrive with a size that is a whole-number multiple of stocks and at a price that is a whole-number multiple of cents. Although this tick size is common to all stocks, the prices of different stocks on Nasdaq vary across several orders of magnitude (from about $\$1$ to more than $\$1000$). Therefore, the \emph{relative tick size} (i.e., the ratio between the stock price and the tick size) varies considerably across different stocks. The results of \citet{Cont:2014price} suggest that the strength of the relationship in equation \eqref{eq:cont_linear_model} may be influenced by a stock's relative tick size. To investigate this possible effect in our own empirical calculations, we choose six stocks with different relative tick sizes: Amazon (AMZN), Tesla (TSLA), Netflix (NFLX), Oracle (ORCL), Cisco (CSCO), and Micron Technology (MU). We provide summary statistics for these stocks in Section \ref{sec:summary_stats}.

The Nasdaq platform operates continuous trading from $09$:$30$ to $16$:$00$ on each weekday. Trading does not occur on weekends or public holidays, so we exclude these days from our analysis. Similarly to \citet{Cont:2014price}, we also exclude all activity during the first and last $30$ minutes of each trading day, to ensure that our results are not affected by the abnormal trading behaviour that can occur shortly after the opening auction or shortly before the closing auction (see also the Appendix). We therefore study all trading activity from $10$:$00$ to $15$:$30$. Our data contains all order-flow activity that occurs within this time period on each trading day from 4 January 2016 to 30 December 2016.

The LOBSTER data contains some information about hidden liquidity. In an LOB, orders reflect market participants' views and trading intentions. The MLOFI vector aggregates order flows to reflect the relative strengths of supply and demand at different price levels. However, a hidden limit order is designed to conceal its owner's intentions from other market participants. Given that hidden limit orders are not observable by other market participants, we choose to exclude all activity related to hidden limit orders from our analysis .

The LOBSTER data has many features that make it particularly suitable for our study. First, the data is recorded directly by the Nasdaq servers. Therefore, we avoid the many difficulties (such as misaligned time stamps or incorrectly ordered events) associated with data sets that are recorded by third-party providers. Second, the data is fully self-consistent, in the sense that it does not contain any activities or updates that would violate the rules of LOB trading. By contrast, many other LOB data sets suffer from recording errors that can constitute a considerable source of noise when performing detailed analysis. Third, each limit order described in the data constitutes a firm commitment to trade. Therefore, our results reflect the market dynamics for real trading opportunities, not ``indicative'' declarations of possible intent.

The LOBSTER data provides several important benefits over the TAQ data used by \cite{Cont:2014price}. First, all market events in LOBSTER are recorded with extremely precise timestamps, whereas the timestamps of market events in TAQ are rounded to the nearest second. This greater time precision gives us greater sampling accuracy for our empirical calculations.\footnote{The LOBSTER data reports timestamps to the accuracy of nanoseconds. Due to the latencies inherent in any computer system, it is unlikely that these timestamps are truly accurate to such extreme precision, but having access to data at this resolution is much better than only having access to data where arrival times are coarse-grained to the nearest second.} Second, all events in LOBSTER are recorded with their exact order size, while the TAQ database only reports round-number lot-size changes. Therefore, we are able to use the exact order size, rather than an approximation, in our calculations. Third, the LOBSTER database provides this detailed information up to any specified number of price levels beyond the bid- and ask-prices. This allows us to calculate the MLOFI vector for any desired choice of $M$.

The LOBSTER database describes all LOB activity that occurs on Nasdaq, but it does not provide any information regarding order flow for the same assets on different platforms. To minimize the possible impact on our results, we restrict our attention to stocks for which Nasdaq is the primary trading venue and therefore captures the majority of order flow. Our results enable us to identify several robust statistical regularities linking MLOFI and mid-price movements, which is precisely the aim of our study. We therefore do not regard this feature of the LOBSTER data to be a serious limitation for our study.

\subsection{Summary Statistics}\label{sec:summary_stats}

Table \ref{tab:lob_means} lists the mean mid-price $\langle P(t)\rangle$, mean bid--ask spread $\langle s(t)\rangle$ and mean number of shares available (i.e., the total size of all limit orders) at the level-1, level-2, level-3, level-4, and level-5 bid- and ask-queues. Among the six stocks, AMZN has the highest mean mid-price $\langle P(t)\rangle$ = \$699.22 while MU has the lowest $\langle P(t)\rangle$ = \$14.16. Therefore, AMZN has the smallest relative tick size and MU has the largest relative tick size.

\begin{table} [!ht]
\begin{center}
\begin{tabular}{|*{7}{P{1.5cm}|}}
\hline
 & \textbf{AMZN} & \textbf{TSLA} & \textbf{NFLX} & \textbf{ORCL} & \textbf{CSCO} & \textbf{MU} \\ 
\hline
$\langle P(t)\rangle$ & 699.22  & 209.81  & 101.98 & 39.22 & 28.78 & 14.16 \\ 
\hline
$\langle s(t)\rangle$ & 0.367   & 0.192   & 0.040 & 0.012 & 0.011 & 0.011 \\ 
\hline
$\langle r^1(t)\rangle $ & 131 & 178 & 288 &  2,386 & 9,406 &  7,255 \\ 
$\langle r^2(t)\rangle $ & 118 & 168 & 344 &  2,766 & 12,001 & 8,291 \\ 
$\langle r^3(t)\rangle $ & 109 & 160 & 384 &  3,012 & 12,789 & 7,817 \\ 
$\langle r^4(t)\rangle $ & 106 & 155 & 411 &  3,007 & 11,347 & 7,751 \\ 
$\langle r^5(t)\rangle $ & 105 & 154 & 429 &  2,501 & 9,945 & 7,426 \\ 
\hline
$\langle q^1(t)\rangle $ & 135 & 175 & 305 &  2,477 & 10,006 & 7,397 \\ 
$\langle q^2(t)\rangle $ & 121 & 173 & 362 &  2,881 & 12,578 & 8,612 \\ 
$\langle q^3(t)\rangle $ & 112 & 170 & 402 &  3,011 & 12,858 & 8,139 \\ 
$\langle q^4(t)\rangle $ & 110 & 168 & 424 &  3,002 & 12,008 & 8,051 \\ 
$\langle q^5(t)\rangle $ & 109 & 168 & 441 &  2,499 & 10,869 & 7,759 \\
\hline
\end{tabular}
\end{center}
\caption{Mean mid-price (measured in US dollars), mean bid--ask spread (measured in US dollars), and mean number of shares available (i.e., the total size of all limit orders) at the level-1, level-2, level-3, level-4, and level-5 bid- and ask-queues (measured in number of shares) for the 6 stocks in our study during the full trading year of 2016.}
\label{tab:lob_means}
\end{table}

As Table \ref{tab:lob_means} illustrates, the larger the relative tick size, the smaller the mean bid-ask spread $\langle s(t) \rangle$ and the larger the mean number of shares available at the level-1 bid- and ask-prices. We observe considerable variation in $\langle s(t) \rangle$ across the stocks in our sample, ranging from about 37 ticks for AMZN to about 1 tick for MU.

Given that we seek to investigate how the order-flow activity at different levels in an LOB influences the mid-price, it is interesting to first understand the similarities and differences between the coarse-grained order flows for each of the six stocks in our sample. Table \ref{tab:spread_quotes_beyond} shows the concentration of order-flow activity (i.e., market order arrivals, limit order arrivals, and cancellations) that occurs within the bid--ask spread, at the bid- or ask-price, and deeper into the LOB, measured as a percentage of the (left panel) total number and (right panel) total volume of all order-flow activity for the given stock.

\begin{table} [!ht]
\begin{center}
\begin{tabular}{|*{7}{P{1.5cm}|}}
\hline
 & \multicolumn{3}{c|}{Number of Orders} & \multicolumn{3}{c|}{Volume of Orders} \\ \hline
 & within bid--ask spread & at bid- or ask-price & deeper into the LOB & within bid--ask spread & at bid- or ask-price & deeper into the LOB \\ \hline
\textbf{AMZN} & 21.31\% & 25.60\% & 53.10\% & 22.92\% & 22.98\% & 53.10\% \\
\textbf{TSLA} & 23.12\% & 27.27\% & 49.61\% & 26.90\% & 25.48\% & 47.62\%  \\
\textbf{NFLX} & 13.81\% & 32.77\% & 53.42\% & 16.39\% & 32.18\% & 51.43\%  \\
\textbf{ORCL} & 1.20\% & 69.05\% & 29.75\% & 1.34\% & 65.34\% & 33.32\%  \\
\textbf{CSCO} & 0.59\% & 68.72\% & 30.69\% & 0.88\% & 68.22\% & 30.90\%  \\
\textbf{MU} & 0.83\% & 70.21\% & 28.97\% & 1.56\% & 72.51\% & 25.93\%  \\
\hline
\end{tabular}
\end{center}
\caption{Percentage of all order-flow activity (i.e., market order arrivals, limit order arrivals, and cancellations) that occurs within the bid--ask spread, at the bid- or ask-price, and deeper into the LOB, measured as a percentage of the (left panel) total number and (right panel) total volume of all order-flow activity for the given stock.}
\label{tab:spread_quotes_beyond}
\end{table}

For the large-tick stocks (i.e., ORCL, CSCO and MU), the majority of order-flow activity occurs at the best bid- or ask-price. Intuitively, this makes sense: The bid--ask spread for such stocks is usually at its minimum possible value of $1$ tick, and new limit orders cannot arrive inside the spread whenever this is the case. This causes limit orders to stack up at the best quotes (see Table \ref{tab:lob_means}). Due to the large number of shares available at these prices, it is relatively unlikely that a single market order will match to limit orders beyond the best quotes. This could explain why so few market participants choose to place limit orders deeper into the LOB.

For the small-tick stocks (i.e., AMZN, TSLA and NFLX), the pattern is quite different. A much larger percentage of order-flow activity occurs inside the bid--ask spread, and a somewhat larger percentage of order-flow activity occurs deeper into the LOB. Even though the relative tick size varies considerably across the small-tick stocks in our sample, in all cases more than half of all order flow occurs beyond the level-1 bid- and ask-prices. Understanding the extent to which such order flow impacts changes in the mid-price is the main focus of our study.

\subsection{Sample Construction}\label{sec:sample_construction}

%We adopt the same sampling methodology as \citet{Cont:2014price} for constructing our data samples. Specifically, for each stock and each trading day, we first divide the trading day into a sequence of consecutive windows, each with the same length. We write $T_i$ to denote  the start time of the $i^{\text{th}}$ window, such that a whole trading day may be partitioned using the (uniformly spaced) grid

To construct our data samples, we adopt the same methodology as \citet{Cont:2014price} (see Section \ref{sec:ofi}). For the results that we show throughout the remainder of the paper, we partition each trading day into $I=11$ uniformly spaced, non-overlapping windows of length $\Delta T = 30$ minutes, then sub-divide each of these windows into $K=180$ uniformly spaced, non-overlapping windows of length $\Delta t = 10$ seconds. For each time interval $\left(t_{i,k-1}, t_{i,k}\right]$, we measure the change in mid-price $\Delta P(t_{i,k-1}, t_{i,k})$ and the $MLOFI^m(t_{i,k-1}, t_{i,k})$ for $m=1,\ldots,M$, then perform the relevant regression fits. For each stock, this provides us with $252 \times 11 =2772$ regression fits. For the fitted regression parameters, we report the mean fitted value and the mean standard error from these $2772$ regressions.

\subsubsection{Window Lengths}

We also repeated all of our calculations with a range of different choices of $I$ and $K$, yielding window lengths that ranged from $\Delta T=30$ minutes to $\Delta T=60$ minutes, and from $\Delta t=5$ seconds to $\Delta t=40$ seconds. For the large-tick stocks in our sample (i.e., ORCL, CSCO, and MU), our results were qualitatively similar for all choices of window lengths in this range. For the small-tick stocks in our sample (i.e., AMZN, TSLA, and NFLX), we found that the adjusted $R^2$ increased slightly as we increased the values of both $\Delta T$ and $\Delta t$. We found the largest variation to occur for AMZN, for which we observed the adjusted $R^2$ to vary by about $10\%$ across this range.

\subsubsection{Intra-Day Seasonalities}\label{sec:intraday}

Both \citet{Cont:2014price} and \citet{Mertens:2019liquidity} reported intra-day seasonalities in the OFI relationships that they studied. To assess the extent to which such intra-day seasonalities may affect our own results, we also repeated all of our OLS regression fits for each of the $I=11$ time windows separately. We discuss our methodology and describe our results for these calculations in the Appendix.

Consistently with both \citet{Cont:2014price} and \citet{Mertens:2019liquidity}, we find that for the largest-tick stocks in our sample (i.e., CSCO and MU), the fitted values of the $\beta^1$ parameters are smaller for the time windows that occur later within the trading day. For all of the other stocks in our sample, we find that the magnitude of this effect is very small, and is difficult to separate from the underlying random fluctuations. We find a similar result for all of the other $\beta^m$ parameters (i.e., $\beta^2$, $\beta^3$, \ldots, $\beta^{10}$) for all of the stocks in our sample. Therefore, given that the magnitude of this effect is small, and given also that intra-day seasonalities are not the core focus of this work, we choose to present all of our empirical calculations throughout the remainder of the paper when averaging across all $I=11$ time windows.

\section{Results}\label{sec:results}

% We now turn our attention to our main contribution: assessing the relationship between MLOFI and changes in the mid-price, across the first $M=10$ price levels. By doing so, we are able to assess the extent to which net order flow at various price levels in the LOB provide explanatory power for the contemporaneous change in mid-price.

\subsection{OLS Fits for OFI}\label{sec:results_ofi}

We first use OLS to estimate the coefficients of the OFI equation \eqref{eq:ofi_regression}. We use these OFI regression fits as a baseline against which to compare our MLOFI regression fits in subsequent sections. By doing so, we are able to quantify the additional explanatory power provided by order-flow imbalance at price levels deeper into the LOB (see Table \ref{table_8}).

Table \ref{tab:ofi_ols} shows the mean fitted regression coefficients and their mean standard errors, each taken across the $2772$ regressions that we perform for each stock. Similarly to \citet{Cont:2014price}, we find that $\alpha \approx 0$ in all cases. This suggests an approximately symmetric behaviour between buying and selling activity. We also find that the mean value of $\beta$ is positive, and several standard deviations greater than 0, for all of the stocks in our sample.

\begin{table} [!ht]
\begin{center}
\begin{tabular}{|c|c|c|c|c|c|c|}
\hline
 & \textbf{AMZN} & \textbf{TSLA} & \textbf{NFLX} & \textbf{ORCL} & \textbf{CSCO} & \textbf{MU} \\ 
\hline
$\alpha$ & $\boldsymbol{0.00}\ (0.83)$ & $\boldsymbol{0.01}\ (0.45)$ & $\boldsymbol{0.00}\ (0.16)$ & $\boldsymbol{0.00}\ (0.02)$ & $\boldsymbol{0.00}\ (0.02)$ & $\boldsymbol{0.00}\ (0.02)$ \\
$\beta$ & $\boldsymbol{11.00}\ (0.92)$ & $\boldsymbol{5.82}\ (0.51)$ & $\boldsymbol{3.11}\ (0.17)$ & $\boldsymbol{0.63}\ (0.02)$ & $\boldsymbol{0.49}\ (0.02)$ & $\boldsymbol{0.56}\ (0.02)$ \\
\hline
\end{tabular}
\caption{OLS estimates of the intercept coefficient $\alpha$ and slope coefficient $\beta$ in the OFI regression equation \eqref{eq:ofi_regression}. The numbers in bold denote the mean value of the fitted parameters and the numbers in parentheses denote the mean of the standard error, each taken across the $2772$ regressions that we perform.}
\label{tab:ofi_ols}
\end{center}
\end{table}

Following \citet{Cont:2014price}, we also calculate the $t$-statistics, $p$-values and percentage of samples that are significant at the $95\%$ level (see Table \ref{tab:ofi_stat_sig}). In all cases, these results indicate that the intercept coefficient is not statistically significant, but that the slope coefficient is strongly statistically significant. Consistently with \citet{Cont:2014price}, we therefore conclude that price movement is an increasing function of OFI. Put simply: The larger the OFI in a given time interval, the larger the expected contemporaneous change in mid-price. 

\begin{table}
\begin{center}
\begin{tabular}{|c|ccc|ccc|ccc|}
\hline
 & \multicolumn{3}{c|}{AMZN} & \multicolumn{3}{c|}{TSLA} & \multicolumn{3}{c|}{NFLX} \\
\hline
 & $t$-stat & $p$-value & count \% & $t$-stat & $p$-value & count \% & $t$-stat & $p$-value & count \% \\ \hline
$\alpha$ & -0.02  & 0.36  & 23\%  & 0.02  & 0.37  & 21\%  & 0.04  & 0.37  & 22\% \\
$\beta$ & 12.20 & $<0.01$  & 100\%  & 11.86  & $<0.01$  & 100\%  & 20.22  & $<0.01$  & 100\% \\
\hline
\hline
 & \multicolumn{3}{c|}{ORCL} & \multicolumn{3}{c|}{CSCO} & \multicolumn{3}{c|}{MU} \\
\hline
 & $t$-stat & $p$-value & count \% & $t$-stat & $p$-value & count \% & $t$-stat & $p$-value & count \% \\ \hline
$\alpha$ & 0.07  & 0.43  & 12\%  & 0.03  & 0.51  & 5\%  & 0.04  & 0.44  & 11\% \\
$\beta$ & 28.39 & $<0.01$  & 100\%  & 29.14  & $<0.01$  & 100\%  & 30.08  & $<0.01$  & 100\% \\
\hline
\end{tabular}
\end{center}
\caption{Statistical significance tests (i.e., mean $t$-statistic, mean $p$-value, and the percentage of samples that are significant at the $95\%$ level) of the intercept coefficient $\alpha$ and slope coefficient $\beta$ in the OFI equation \eqref{eq:ofi_regression}, taken across the $2772$ regressions that we perform.}
\label{tab:ofi_stat_sig}
\end{table}

\subsection{OLS Fits for MLOFI}\label{sec:linear_MLOFI}

We now use OLS to fit the linear model
\begin{equation}\label{eq:mlofi_regression}
\Delta P(t_{i,k-1}, t_{i,k}) = \alpha + \sum^M_{m=1} \beta^m MLOFI^m(t_{i,k-1}, t_{i,k}) + \varepsilon,
\end{equation}
for $M=10$. In this model, the larger the price-impact coefficient $\beta^m$, the more strongly the $m^{\text{th}}$ component of the $MLOFI$ vector contributes to the price change $\Delta P(t_{i,k-1}, t_{i,k})$. Therefore, we can interpret the coefficient $\beta^m$ as a measure of the relative contribution to the price change from the level-$m$ net order-flow imbalance.

Table \ref{tab:mlofi_ols} shows the mean fitted regression coefficients and their mean standard errors, each taken across the $2772$ regressions that we perform for each stock. Table \ref{tab:ols_tests} shows the $t$-statistics, $p$-values and percentage of samples that are significant at the $95\%$ level.

\begin{table} [!ht]
\begin{center}
\begin{tabular}{|c|c|c|c|c|c|c|}
\hline
 & \textbf{AMZN} & \textbf{TSLA} & \textbf{NFLX} & \textbf{ORCL} & \textbf{CSCO} & \textbf{MU} \\ 
\hline
$\alpha$ & $\boldsymbol{-0.05}\ (0.66)$ & $\boldsymbol{0.01}\ (0.36)$ & $\boldsymbol{0.00}\ (0.11)$ & $\boldsymbol{0.00}\ (0.01)$ & $\boldsymbol{0.01}\ (<0.01)$ & $\boldsymbol{0.00}\ (<0.01)$ \\
$\beta^1$ & $\boldsymbol{3.16}\ (1.41)$ & $\boldsymbol{1.94}\ (0.76)$ & $\boldsymbol{0.59}\ (0.26)$ & $\boldsymbol{0.04}\ (0.02)$ & $\boldsymbol{0.02}\ (0.01)$ & $\boldsymbol{0.03}\ (0.01)$ \\
$\beta^2$ & $\boldsymbol{2.49}\ (1.85)$ & $\boldsymbol{1.29}\ (1.01)$ & $\boldsymbol{0.43}\ (0.36)$ & $\boldsymbol{0.06}\ (0.02)$ & $\boldsymbol{0.04}\ (0.01)$ & $\boldsymbol{0.04}\ (0.01)$ \\
$\beta^3$ & $\boldsymbol{2.52}\ (2.09)$ & $\boldsymbol{1.15}\ (1.12)$ & $\boldsymbol{0.40}\ (0.40)$ & $\boldsymbol{0.05}\ (0.02)$ & $\boldsymbol{0.03}\ (0.01)$ & $\boldsymbol{0.06}\ (0.02)$ \\
$\beta^4$ & $\boldsymbol{1.47}\ (2.21)$ & $\boldsymbol{0.83}\ (1.23)$ & $\boldsymbol{0.41}\ (0.44)$ & $\boldsymbol{0.05}\ (0.02)$ & $\boldsymbol{0.05}\ (0.02)$ & $\boldsymbol{0.08}\ (0.02)$ \\
$\beta^5$ & $\boldsymbol{1.13}\ (2.33)$ & $\boldsymbol{0.74}\ (1.30)$ & $\boldsymbol{0.34}\ (0.47)$ & $\boldsymbol{0.07}\ (0.02)$ & $\boldsymbol{0.06}\ (0.02)$ & $\boldsymbol{0.08}\ (0.02)$ \\
$\beta^6$ & $\boldsymbol{1.07}\ (2.44)$ & $\boldsymbol{0.79}\ (1.37)$ & $\boldsymbol{0.28}\ (0.48)$ & $\boldsymbol{0.09}\ (0.03)$ & $\boldsymbol{0.09}\ (0.02)$ & $\boldsymbol{0.10}\ (0.02)$ \\
$\beta^7$ & $\boldsymbol{0.98}\ (2.54)$ & $\boldsymbol{0.70}\ (1.43)$ & $\boldsymbol{0.36}\ (0.49)$ & $\boldsymbol{0.09}\ (0.03)$ & $\boldsymbol{0.09}\ (0.02)$ & $\boldsymbol{0.08}\ (0.02)$ \\
$\beta^8$ & $\boldsymbol{0.75}\ (2.59)$ & $\boldsymbol{0.62}\ (1.45)$ & $\boldsymbol{0.24}\ (0.50)$ & $\boldsymbol{0.08}\ (0.03)$ & $\boldsymbol{0.08}\ (0.02)$ & $\boldsymbol{0.08}\ (0.02)$ \\
$\beta^9$ & $\boldsymbol{0.79}\ (2.63)$ & $\boldsymbol{0.41}\ (1.46)$ & $\boldsymbol{0.30}\ (0.50)$ & $\boldsymbol{0.07}\ (0.03)$ & $\boldsymbol{0.07}\ (0.02)$ & $\boldsymbol{0.07}\ (0.02)$ \\
$\beta^{10}$ & $\boldsymbol{1.09}\ (2.05)$ & $\boldsymbol{0.71}\ (1.18)$ & $\boldsymbol{0.76}\ (0.39)$ & $\boldsymbol{0.10}\ (0.02)$ & $\boldsymbol{0.06}\ (0.02)$ & $\boldsymbol{0.08}\ (0.02)$ \\
\hline
\end{tabular}
\caption{OLS estimates of the coefficients in the MLOFI regression equation \eqref{eq:mlofi_regression}. The numbers in bold denote the mean value of the fitted parameters and the numbers in parentheses denote the mean of the standard error, each taken across the $2772$ regressions that we perform.}
\label{tab:mlofi_ols}
\end{center}
\end{table}

\begin{table} [H]
\begin{center}
\scalebox{0.75}{
\begin{tabular}{|*{10}{P{1.5cm}|}}
\hline & \multicolumn{3}{c|}{\textbf{AMZN}} & \multicolumn{3}{c|}{\textbf{TSLA}} & \multicolumn{3}{c|}{\textbf{NFLX}} \\
\hline
Ridge & $t$-stat & $p$-value & count \% & $t$-stat & $p$-value & count \% & $t$-stat & $p$-value & count \% \\
\hline
\textbf{$\alpha$} & $-0.07$ & $0.42$ & $16\%$ & $0.02$ & $0.41$ & $16\%$ & $0.09$ & $0.36$ & $22\%$ \\
\textbf{$\beta^1$} & $2.43$ & $0.15$ & $60\%$ & $2.71$ & $0.13$ & $65\%$ & $2.37$ & $0.15$ & $58\%$ \\
\textbf{$\beta^2$} & $1.35$ & $0.28$ & $32\% $ & $1.34$ & $0.29$ & $33\%$ & $1.49$ & $0.26$ & $37\%$ \\
\textbf{$\beta^3$} & $1.23$ & $0.30$ & $30\%$ & $1.09$ & $0.33$ & $28\%$ & $1.32$ & $0.28$ & $33\%$ \\
\textbf{$\beta^4$} & $0.73$ & $0.38$ & $19\%$ & $0.71$ & $0.38$ & $18\%$ & $1.23$ & $0.30$ & $30\%$ \\
\textbf{$\beta^5$} & $0.53$ & $0.40$ & $15\%$ & $0.58$ & $0.40$ & $15\%$ & $0.98$ & $0.34$ & $24\%$ \\
\textbf{$\beta^6$} & $0.46$ & $0.43$ & $13\%$ & $0.63$ & $0.39$ & $17\%$ & $0.86$ & $0.36$ & $22\%$ \\
\textbf{$\beta^7$} & $0.41$ & $0.43$ & $13\%$ & $0.53$ & $0.40$ & $15\%$ & $0.95$ & $0.35$ & $23\%$ \\
\textbf{$\beta^8$} & $0.33$ & $0.43$ & $12\%$ & $0.42$ & $0.41$ & $13\%$ & $0.78$ & $0.37$ & $19\%$ \\
\textbf{$\beta^9$} & $0.31$ & $0.45$ & $11\%$ & $0.37$ & $0.43$ & $13\%$ & $0.76$ & $0.38$ & $19\%$ \\
\textbf{$\beta^{10}$} & $0.63$ & $0.39$ & $18\%$ & $0.74$ & $0.37$ & $21\%$ & $2.04$ & $0.19$ & $48\%$ \\
\hline
\hline & \multicolumn{3}{c|}{\textbf{ORCL}} & \multicolumn{3}{c|}{\textbf{CSCO}} & \multicolumn{3}{c|}{\textbf{MU}} \\
\hline
Ridge & $t$-stat & $p$-value & count \% & $t$-stat & $p$-value & count \% & $t$-stat & $p$-value & count \% \\
\hline
\textbf{$\alpha$} & $0.25$ & $0.22$ & $49\%$ & $0.30$ & $0.16$ & $63\%$ & $-0.18$ & $0.17$ & $59\%$ \\
\textbf{$\beta^1$} & $2.83$ & $0.14$ & $64\%$ & $2.38$ & $0.17$ & $58\%$ & $2.28$ & $0.18$ & $56\%$ \\
\textbf{$\beta^2$} & $3.11$ & $0.11$ & $71\%$ & $2.65$ & $0.15$ & $62\%$ & $2.33$ & $0.17$ & $56\%$ \\
\textbf{$\beta^3$} & $2.55$ & $0.16$ & $59\%$ & $2.20$ & $0.22$ & $50\%$ & $3.66$ & $0.11$ & $74\%$ \\
\textbf{$\beta^4$} & $2.45$ & $0.18$ & $57\%$ & $2.88$ & $0.14$ & $64\%$ & $4.43$ & $0.08$ & $79\%$ \\
\textbf{$\beta^5$} & $3.25$ & $0.11$ & $70\%$ & $3.42$ & $0.12$ & $71\%$ & $4.45$ & $0.09$ & $77\%$ \\
\textbf{$\beta^6$} & $4.05$ & $0.07$ & $81\%$ & $5.82$ & $0.05$ & $86\%$ & $5.71$ & $0.07$ & $83\%$ \\
\textbf{$\beta^7$} & $4.07$ & $0.08$ & $78\%$ & $7.47$ & $0.03$ & $91\%$ & $5.23$ & $0.07$ & $81\%$ \\
\textbf{$\beta^8$} & $3.60$ & $0.10$ & $73\%$ & $6.65$ & $0.06$ & $85\%$ & $5.15$ & $0.08$ & $80\%$ \\
\textbf{$\beta^9$} & $3.08$ & $0.13$ & $65\%$ & $4.84$ & $0.08$ & $79\%$ & $4.38$ & $0.11$ & $75\%$ \\
\textbf{$\beta^{10}$} & $4.56$ & $0.07$ & $83\%$ & $4.29$ & $0.10$ & $77\%$ & $4.72$ & $0.09$ & $78\%$ \\
\hline
\end{tabular}}
\end{center}
\caption{Statistical significance tests (i.e., mean $t$-statistic, mean $p$-value, and the percentage of samples that are significant at the $95\%$ level) of the OLS regression parameter fits for the MLOFI regression equation \eqref{eq:mlofi_regression}, taken across the $2772$ regressions that we perform (see Section \ref{sec:sample_construction}).}
\label{tab:ols_tests}
\end{table}

For the small-tick stocks (i.e., AMZN, TSLA, and NFLX), the $p$-values for all of the fitted $\beta^m$ coefficients are relatively weak. For the large-tick stocks (i.e., ORCL, CSCO, and MU), the $p$-values are weak for the $\beta^m$ coefficients close to the bid--ask spread, but are stronger for values of $m$ greater than about 5. In this range, however, the fitted values of the $\beta^m$ coefficients are all quite close to 0.

When assessed together with our results for OFI (see Tables \ref{tab:ofi_ols} and \ref{tab:ofi_stat_sig}), the results in Table \ref{tab:ols_tests} paint a rather complex picture. In our OLS regression fits of the OFI equation \eqref{eq:ofi_regression}, net order flow at the level-1 bid- and ask-prices is strongly statistically significant (as captured by the parameter $\beta$ in Table \ref{tab:ofi_stat_sig}). In our OLS regression fits of the MLOFI equation \eqref{eq:mlofi_regression}, the same effect from net order flow at the level-1 bid- and ask-prices is much less strong (as captured by the parameter $\beta_1$ in Table \ref{tab:ols_tests}). Given that the net order flow at the level-1 bid- and ask-prices appears as an input in the MLOFI regressions (it is precisely the first component of the MLOFI vector), this result is puzzling.

% Looking to the fitted values of the other $\beta^m$ coefficients does not do much to clarify this picture. For the small-tick stocks (i.e., AMZN, TSLA, and NFLX), none of the other $\beta^m$ coefficients are strongly statistically significant. For the large-tick stocks (i.e., ORCL, CSCO, and MU), the $\beta^m$ coefficients are statistically significant for values of $m$ greater than about 5, but the fitted values of these coefficients are all close to 0. For smaller values of $m$, the $\beta^m$ coefficients are weakly significant at best.

Given that many of the fitted values of $\beta^m$ are not statistically significant, we do not perform a detailed quantitative analysis of these coefficients, because such an analysis could be misleading. Instead, we turn our attention to the more general question: Why do the multiple regressions behave in this way, given that the corresponding results for the OFI regressions are so simple? As we discuss in the next section, the answer to this puzzle lies less in the subtlety of LOB price formation than it does in the (un-)suitability of multivariate OLS regression as a tool for performing this analysis.

\subsection{MLOFI Sample Correlations}\label{sec:sample_correlations}

When performing multivariate OLS regression, a key assumption is that the feature variables are all mutually linearly independent. If this assumption does not hold, then the regression matrix is singular, and it is not possible to calculate its inverse to solve the regression equation. If the feature variables are highly correlated, although the regression matrix is technically invertible, the resulting least-squares estimates will be unstable. Unstable OLS regression fits produce parameter estimates that are extremely sensitive to small changes in the input data, and whose variance is so large that their fitted values may be very far from the true values. This phenomenon is called \emph{multicollinearity}.

\begin{figure}[H] 
	\centering
	\includegraphics[width=0.95\textwidth]{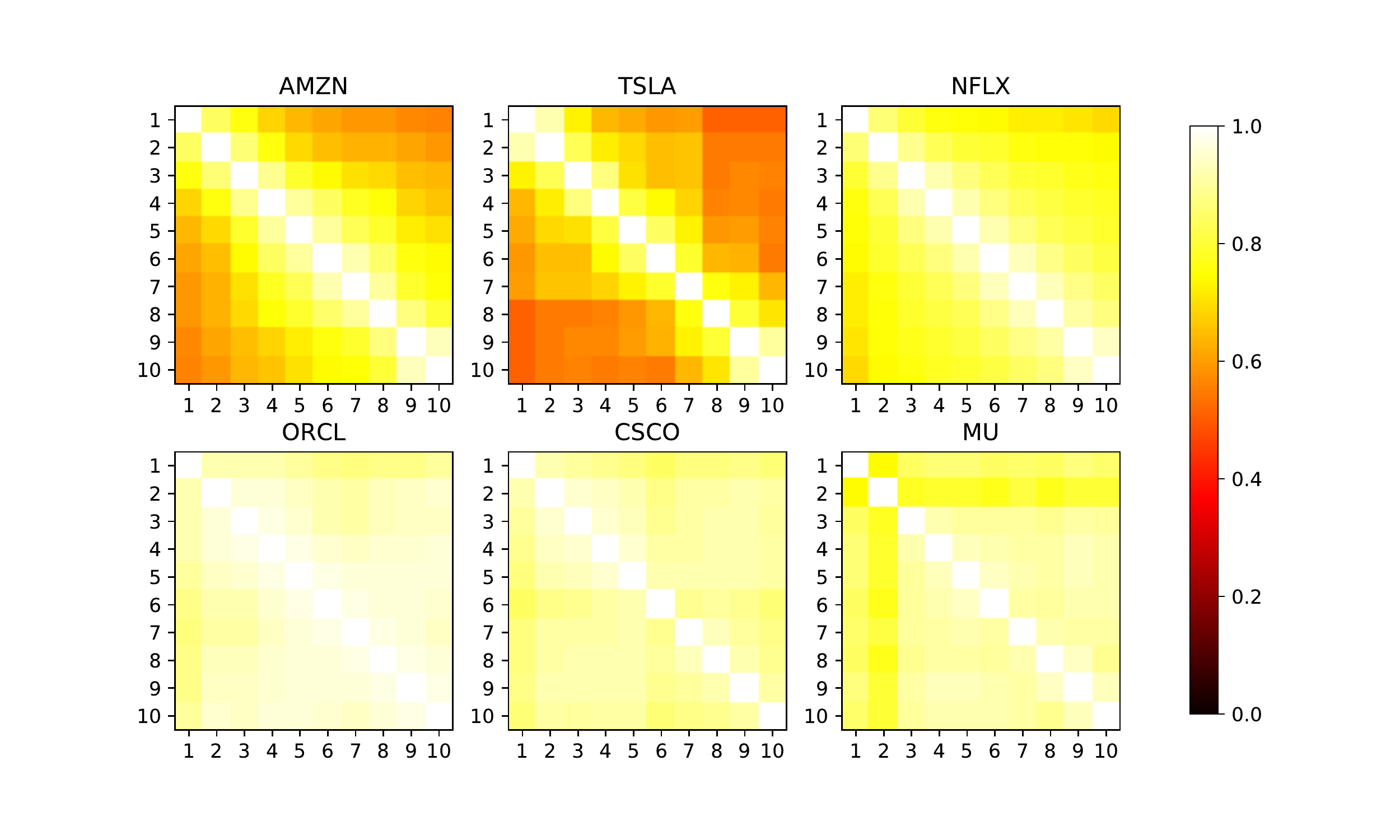}
    \caption{Sample correlations between the $M=10$ components in the MLOFI vectors.}
    \label{fig:sample_correlation}
\end{figure}

To help understand our multivariate OLS regression results in Section \ref{sec:linear_MLOFI}, we now turn our attention to the question of whether or not the feature variables in the MLOFI regression equation \eqref{eq:mlofi_regression} are mutually linearly independent. Given that these feature variables correspond to order-flow imbalance at neighbouring price levels within the same LOB, and given that some order-flow activities affect the values of MLOFI at several different price levels (see the example in Section \ref{sec:mlofi_example}), it is reasonable to expect that our feature variables $MLOFI^m(t_{i,k-1}, t_{i,k})$ may exhibit multicollinearity. To assess whether this is indeed the case, we first calculate the sample correlation matrix for each stock (see Figure \ref{fig:sample_correlation}).

In all cases, the sample correlation matrices reveal strong multicollinearity. For the small-tick stocks in our sample (i.e., AMZN, TSLA, and NFLX), the sample correlation decreases somewhat with increasing distance between price levels, but it still remains above $0.5$ even when comparing the first and tenth components of the MLOFI vectors. For all other stocks in our sample, the sample correlation is above $0.7$ for all pairs of components in the MLOFI vector.

\begin{figure}[H] 
	\centering
	\includegraphics[width=0.7\textwidth]{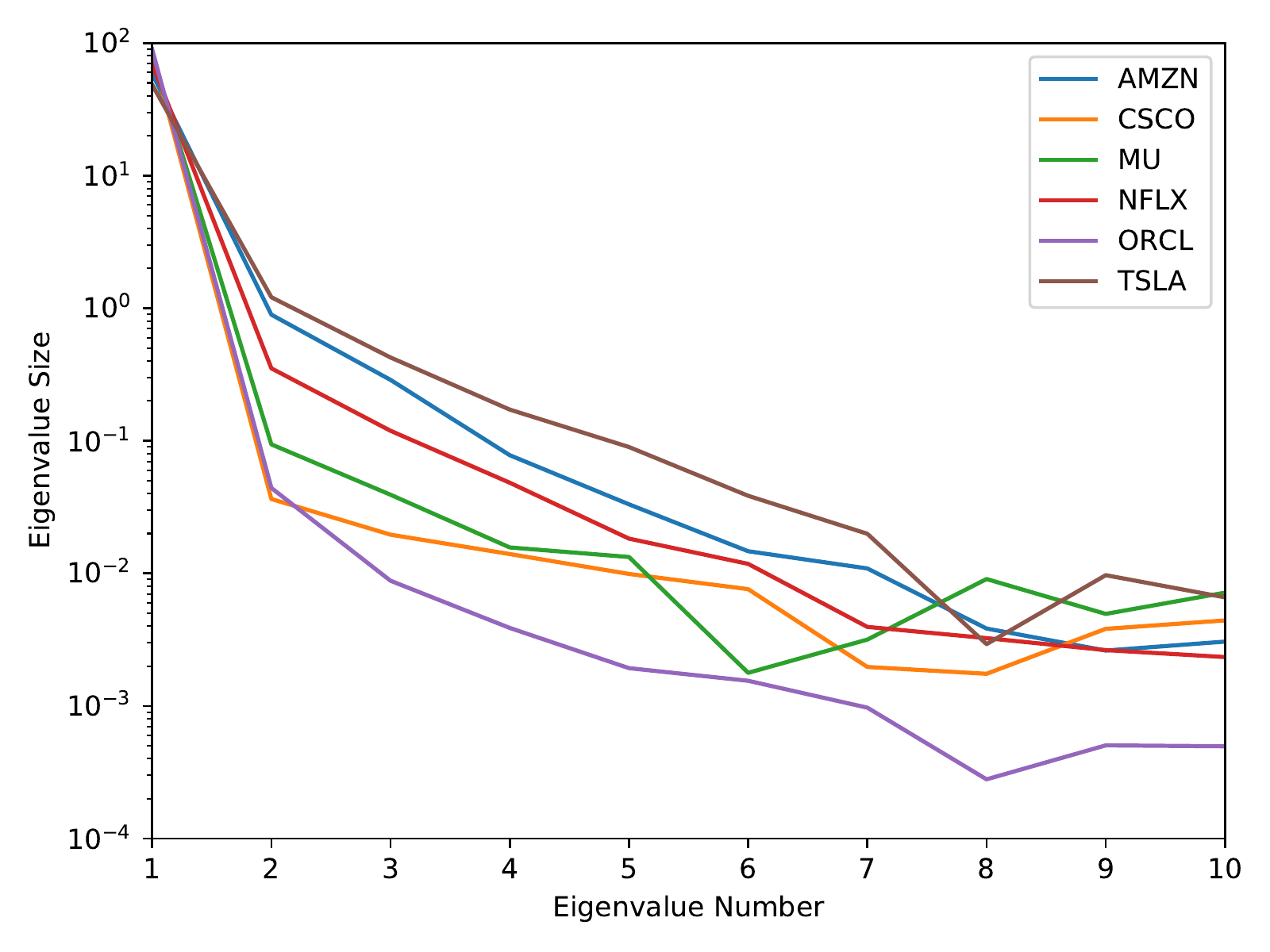}
    \caption{Eigenvalues of the sample correlation matrices shown in Figure \ref{fig:sample_correlation}.}
    \label{fig:eigenvalues}
\end{figure}

To assess the extent to which this may cause instability in the resulting OLS regression estimates, we also calculate the corresponding eigenvalues (see Figure \ref{fig:eigenvalues}). The eigenvalues are typically larger for the small-tick stocks (i.e., AMZN, TSLA, and NFLX) than they are for the large-tick stocks (i.e., ORCL, CSCO, and MU), but in all cases, and for all $i\geq 2$, the ratio of the $i^\text{th}$ eigenvalue to the first eigenvalue is always very close to 0. This strongly suggests that the parameter estimates obtained by multivariate OLS regression are unstable. Therefore, direct interpretation of the $\beta^m$ coefficients fitted by OLS regression may produce a misleading picture of how net order flow at the level-$m$ bid- and ask-prices truly impacts the contemporaneous change in mid-price.

\subsection{Ridge Regression Fits for MLOFI}\label{sec:results_mlofi}

To address the problem of multicollinearity that we uncovered in Section \ref{sec:sample_correlations}, we now pursue an alternative methodology for fitting the parameters of the MLOFI equation \eqref{eq:mlofi_regression}. Specifically, we implement \emph{Ridge regression,} which seeks to to combat the effects of multicollinearity by introducing a regularization term into the multiple regression equation. For a full discussion of Ridge regression, see \citet{hoerl1970ridge}.

For a given stock during a given time window of length $\Delta T$, and for a given choice of $M$, let

\begin{align*}
\boldsymbol{y} &:= (\Delta P_1,\cdots,\Delta P_K)^T, \\
X &:=
\begin{bmatrix}
    1       & MLOFI^1_1 & \dots & MLOFI^M_1 \\
    \vdots  & \vdots & \ddots & \vdots \\
    1       & MLOFI^1_K & \dots & MLOFI^M_K
\end{bmatrix}, \\
\boldsymbol{\beta} &:= (\alpha, \beta^1,\cdots,\beta^M)^T, \text{ and} \\
\boldsymbol{\varepsilon} &:= (\varepsilon_1,\cdots,\varepsilon_K)^T,
\end{align*}
where the elements of $\boldsymbol{\varepsilon}$ are independent and identically distributed (iid) Gaussian random variables with mean $0$. Using this notation, we can rewrite the MLOFI linear model as
\begin{equation}\label{eq:mlofi_general_matrix}
\boldsymbol{y} = X \boldsymbol{\beta} + \boldsymbol{\varepsilon}.
\end{equation}

In Ridge regression, rather than simply seeking the values of $\boldsymbol{\beta}$ that minimize the least-squares cost function $||\boldsymbol{y}-X\boldsymbol{\beta} ||_2^2$, we instead seek the values of $\boldsymbol{\beta}$ that minimize the \emph{Ridge regression cost function}
\begin{equation}\label{eq:ridge_regression}
C(\boldsymbol{\beta},\lambda)=||\boldsymbol{y}-X\boldsymbol{\beta} ||_2^2  + \lambda||\boldsymbol{\beta}  ||_2^2,
\end{equation}
where the hyperparameter $\lambda \geq 0$ controls the strength of the regularization. Intuitively, for any choice of $\lambda$, the term $\lambda||\boldsymbol{\beta}  ||_2^2$ operates as a penalty term in the cost function. The larger the magnitude of the regression parameters, the larger the penalty. Unstable OLS regression fits typically lead to fitted regression parameters with extremely large magnitude. In Ridge regression, the regularization helps to move the global maximum of the cost function away from an otherwise unstable regression fit.

\subsubsection{Choosing the Regularization Parameter $\lambda$}\label{sec:cross_validation}

We use $5$-fold cross validation to choose a suitable value of the regularization parameter $\lambda$. Specifically, we consider a range of 50 logaritmically-spaced candidate values of $\lambda$ between $10^{-5}$ and $10^5$, and we choose the value $\hat{\lambda}$ that yields the smallest cross-validation error. For a detailed introduction to cross-validation, see \citet{Hastie:2009elements}.

\begin{figure}[H]
	\centering
	\includegraphics[width=0.7\textwidth]{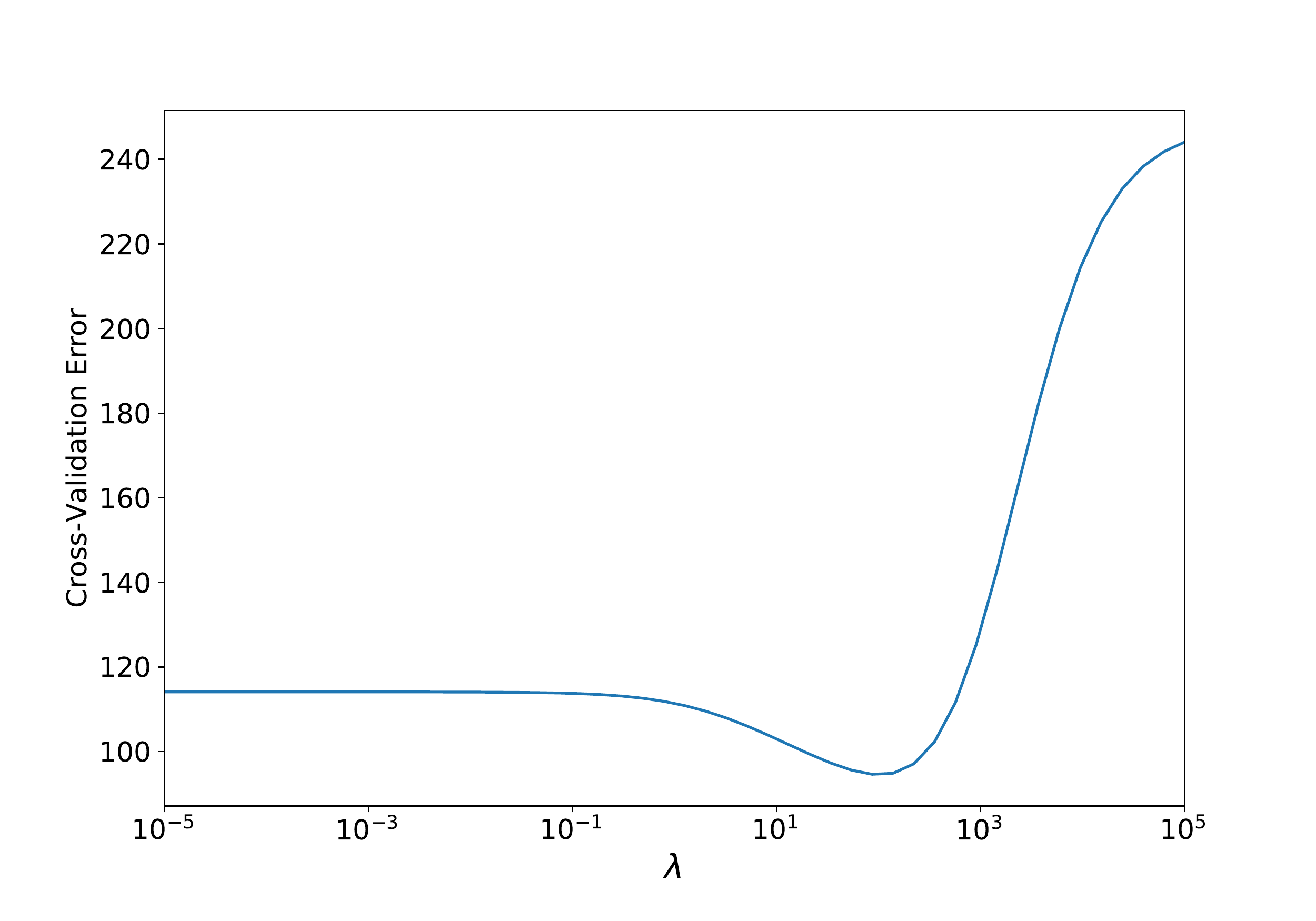}
    \caption{Mean cross-validation error AMZN. We consider a range of 50 logarithmically-spaced candidate values of $\lambda$ between $10^{-5}$ and $10^5$.}
    \label{fig:Lambda}
\end{figure}

To illustrate this process, Figure \ref{fig:Lambda} shows the mean cross-validation error as a function of $\lambda$ for AMZN. As the figure shows, there is a clear local minimum in the mean cross-validation error, at $\hat{\lambda}\approx 139$. Table \ref{tab:cv_results} shows that corresponding values of $\hat{\lambda}$ for each of the stocks in our sample.

\begin{table} [!ht]
\begin{center}
\begin{tabular}{|c|c|c|c|c|c|c|}
\hline
 & \textbf{AMZN} & \textbf{TSLA} & \textbf{NFLX} & \textbf{ORCL} & \textbf{CSCO} & \textbf{MU} \\ 
\hline
$\hat{\lambda}$ & $139.0$ & $54.3$ & $33.9$ & $3.2$ & $1.3$ & $2.0$ \\
\hline
\end{tabular}
\end{center}
\caption{Cross-validation estimates of $\hat{\lambda}$ in the Ridge regression cost function \eqref{eq:ridge_regression}. We consider a range of 50 logaritmically-spaced candidate values of $\lambda$ between $10^{-5}$ and $10^5$ (see Figure \ref{fig:Lambda}).}
\label{tab:cv_results}
\end{table}

\subsubsection{Ridge Regression Fits for MLOFI}

Table \ref{tab:ridge_results} shows the mean fitted regression coefficients and their mean standard errors,\footnote{We obtain the standard errors for Ridge regression by the formula given in Section 1.4.2 of \citet{van2015lecture}.} each taken across the $2772$ regressions that we perform for each stock. Table \ref{tab:ridge_tests} shows the corresponding $t$-statistics, $p$-values and percentage of samples that are significant at the $95\%$ level.

\begin{table} [H]
\begin{center}
\scalebox{0.75}{
% \begin{tabular}{|*{7}{P{2.5cm}|}}
\begin{tabular}{|c|c|c|c|c|c|c|c|c|c|c|c|c|}
\hline & \multicolumn{2}{c|}{\textbf{AMZN}} & \multicolumn{2}{c|}{\textbf{TSLA}} & \multicolumn{2}{c|}{\textbf{NFLX}} & \multicolumn{2}{c|}{\textbf{ORCL}} & \multicolumn{2}{c|}{\textbf{CSCO}} & \multicolumn{2}{c|}{\textbf{MU}} \\
\hline
\textbf{$\alpha$} & $\boldsymbol{-0.05}$ & $(0.43)$ & $\boldsymbol{0.01}$ & $(0.23)$ & $\boldsymbol{0.01}$ & $(0.07)$ & $\boldsymbol{0.00}$ & $(0.01)$ & $\boldsymbol{0.00}$ & $(<0.01)$ & $\boldsymbol{0.00}$ & $(<0.01)$ \\
\textbf{$\beta^1$} & $\boldsymbol{2.17}$ & $(0.50)$ & $\boldsymbol{1.28}$ & $(0.26)$ & $\boldsymbol{0.46}$ & $(0.10)$ & $\boldsymbol{0.05}$ & $(0.01)$ & $\boldsymbol{0.03}$ & $(0.01)$ & $\boldsymbol{0.04}$ & $(0.01)$ \\
\textbf{$\beta^2$} & $\boldsymbol{1.99}$ & $(0.49)$ & $\boldsymbol{1.04}$ & $(0.26)$ & $\boldsymbol{0.42}$ & $(0.10)$ & $\boldsymbol{0.06}$ & $(0.01)$ & $\boldsymbol{0.04}$ & $(0.01)$ & $\boldsymbol{0.04}$ & $(0.01)$ \\
\textbf{$\beta^3$} & $\boldsymbol{1.85}$ & $(0.49)$ & $\boldsymbol{0.90}$ & $(0.25)$ & $\boldsymbol{0.39}$ & $(0.10)$ & $\boldsymbol{0.05}$ & $(0.01)$ & $\boldsymbol{0.03}$ & $(0.01)$ & $\boldsymbol{0.06}$ & $(0.01)$ \\
\textbf{$\beta^4$} & $\boldsymbol{1.44}$ & $(0.48)$ & $\boldsymbol{0.78}$ & $(0.25)$ & $\boldsymbol{0.37}$ & $(0.10)$ & $\boldsymbol{0.05}$ & $(0.01)$ & $\boldsymbol{0.05}$ & $(0.01)$ & $\boldsymbol{0.08}$ & $(0.01)$ \\
\textbf{$\beta^5$} & $\boldsymbol{1.21}$ & $(0.48)$ & $\boldsymbol{0.70}$ & $(0.25)$ & $\boldsymbol{0.34}$ & $(0.10)$ & $\boldsymbol{0.07}$ & $(0.01)$ & $\boldsymbol{0.06}$ & $(0.01)$ & $\boldsymbol{0.08}$ & $(0.01)$ \\
\textbf{$\beta^6$} & $\boldsymbol{1.09}$ & $(0.47)$ & $\boldsymbol{0.69}$ & $(0.25)$ & $\boldsymbol{0.32}$ & $(0.10)$ & $\boldsymbol{0.09}$ & $(0.01)$ & $\boldsymbol{0.08}$ & $(0.01)$ & $\boldsymbol{0.09}$ & $(0.01)$ \\
\textbf{$\beta^7$} & $\boldsymbol{1.01}$ & $(0.47)$ & $\boldsymbol{0.63}$ & $(0.25)$ & $\boldsymbol{0.33}$ & $(0.10)$ & $\boldsymbol{0.09}$ & $(0.01)$ & $\boldsymbol{0.08}$ & $(0.01)$ & $\boldsymbol{0.08}$ & $(0.01)$ \\
\textbf{$\beta^8$} & $\boldsymbol{0.92}$ & $(0.46)$ & $\boldsymbol{0.57}$ & $(0.24)$ & $\boldsymbol{0.32}$ & $(0.10)$ & $\boldsymbol{0.08}$ & $(0.01)$ & $\boldsymbol{0.08}$ & $(0.01)$ & $\boldsymbol{0.08}$ & $(0.01)$ \\
\textbf{$\beta^9$} & $\boldsymbol{0.89}$ & $(0.46)$ & $\boldsymbol{0.53}$ & $(0.25)$ & $\boldsymbol{0.36}$ & $(0.10)$ & $\boldsymbol{0.07}$ & $(0.01)$ & $\boldsymbol{0.07}$ & $(0.01)$ & $\boldsymbol{0.07}$ & $(0.01)$ \\
\textbf{$\beta^{10}$} & $\boldsymbol{1.01}$ & $(0.48)$ & $\boldsymbol{0.60}$ & $(0.25)$ & $\boldsymbol{0.46}$ & $(0.10)$ & $\boldsymbol{0.09}$ & $(0.01)$ & $\boldsymbol{0.06}$ & $(0.01)$ & $\boldsymbol{0.07}$ & $(0.01)$ \\
\hline
\end{tabular}}
\end{center}
\caption{Ridge regression parameter estimates for the MLOFI equation \eqref{eq:mlofi_regression}. The numbers in bold denote the mean value of the fitted parameters and the numbers in parentheses denote the mean of the standard error, each taken across the $2772$ regressions that we perform (see Section \ref{sec:sample_construction}). For each stock, we use $5$-fold cross validation to choose the value of $\lambda$ (see Section \ref{sec:cross_validation}).}
\label{tab:ridge_results}
\end{table}

\begin{table} [H]
\begin{center}
\scalebox{0.75}{
\begin{tabular}{|*{10}{P{1.5cm}|}}
\hline & \multicolumn{3}{c|}{\textbf{AMZN}} & \multicolumn{3}{c|}{\textbf{TSLA}} & \multicolumn{3}{c|}{\textbf{NFLX}} \\
\hline
Ridge & $t$-stat & $p$-value & count \% & $t$-stat & $p$-value & count \% & $t$-stat & $p$-value & count \% \\
\hline
\textbf{$\alpha$} & -0.17 & 0.23 & 29\% & 0.07 & 0.31 & 32\%& 0.16 & 0.29 & 9\% \\
\textbf{$\beta^1$} & 4.87 & 0.02 & 93\% & 5.35 & 0.02 & 94\%& 5.65 & 0.02 & 79\% \\
\textbf{$\beta^2$} & 4.73 & 0.03 & 92\% & 4.90 & 0.03 & 91\%& 5.81 & 0.03 & 78\% \\
\textbf{$\beta^3$} & 4.47 & 0.04 & 89\% & 4.41 & 0.05 & 86\%& 5.87 & 0.04 & 77\% \\
\textbf{$\beta^4$} & 3.92 & 0.07 & 82\% & 4.03 & 0.07 & 81\%& 5.86 & 0.04 & 75\% \\
\textbf{$\beta^5$} & 3.52 & 0.10 & 74\% & 3.79 & 0.09 & 77\%& 5.56 & 0.05 & 72\% \\
\textbf{$\beta^6$} & 3.29 & 0.11 & 70\% & 3.76 & 0.09 & 77\%& 5.31 & 0.06 & 68\% \\
\textbf{$\beta^7$} & 3.14 & 0.13 & 67\% & 3.55 & 0.10 & 73\%& 5.24 & 0.06 & 67\% \\
\textbf{$\beta^8$} & 2.98 & 0.15 & 64\% & 3.30 & 0.12 & 69\%& 5.20 & 0.06 & 67\% \\
\textbf{$\beta^9$} & 2.89 & 0.16 & 63\% & 3.13 & 0.13 & 68\%& 5.37 & 0.05 & 71\% \\
\textbf{$\beta^{10}$} & 2.81 & 0.15 & 64\% & 3.07 & 0.13 & 68\%& 5.76 & 0.03 & 78\% \\
\hline
\hline & \multicolumn{3}{c|}{\textbf{ORCL}} & \multicolumn{3}{c|}{\textbf{CSCO}} & \multicolumn{3}{c|}{\textbf{MU}} \\
\hline
Ridge & $t$-stat & $p$-value & count \% & $t$-stat & $p$-value & count \% & $t$-stat & $p$-value & count \% \\
\hline
\textbf{$\alpha$} & 0.26 & 0.20 & 51\% & 0.27 & 0.16 & 63\%& -0.17 & 0.16 & 60\% \\
\textbf{$\beta^1$} & 5.81 & 0.04 & 89\% & 4.93 & 0.08 & 80\%& 5.86 & 0.07 & 83\% \\
\textbf{$\beta^2$} & 6.75 & 0.03 & 90\% & 6.57 & 0.06 & 84\%& 6.42 & 0.06 & 85\% \\
\textbf{$\beta^3$} & 5.86 & 0.06 & 84\% & 5.69 & 0.12 & 72\%& 8.91 & 0.03 & 92\% \\
\textbf{$\beta^4$} & 6.10 & 0.07 & 83\% & 8.23 & 0.05 & 86\%& 10.59 & 0.02 & 94\% \\
\textbf{$\beta^5$} & 8.07 & 0.03 & 93\% & 9.42 & 0.03 & 92\%& 10.88 & 0.02 & 95\% \\
\textbf{$\beta^6$} & 9.54 & 0.01 & 97\% & 11.39 & 0.01 & 97\%& 11.52 & 0.01 & 97\% \\
\textbf{$\beta^7$} & 9.61 & 0.01 & 97\% & 12.29 & 0.01 & 98\%& 10.80 & 0.01 & 96\% \\
\textbf{$\beta^8$} & 8.94 & 0.02 & 95\% & 11.20 & 0.02 & 96\%& 10.45 & 0.02 & 96\% \\
\textbf{$\beta^9$} & 8.47 & 0.02 & 94\% & 9.53 & 0.02 & 94\%& 9.23 & 0.02 & 93\% \\
\textbf{$\beta^{10}$} & 9.32 & 0.01 & 97\% & 8.51 & 0.03 & 92\%& 9.11 & 0.02 & 94\% \\
\hline
\end{tabular}}
\end{center}
\caption{Statistical significance tests (i.e., mean $t$-statistic, mean $p$-value, and the percentage of samples that are significant at the $95\%$ level) of the Ridge regression parameter fits for the MLOFI regression equation \eqref{eq:mlofi_regression}, taken across the $2772$ regressions that we perform (see Section \ref{sec:sample_construction}).}
\label{tab:ridge_tests}
\end{table}

For the intercept coefficient $\alpha$, our results using Ridge regression are similar to those that we obtain when using OLS regression (see Table \ref{tab:mlofi_ols}). In all cases, we again find that $\alpha \approx 0$, which suggests an approximately symmetric behaviour between the buy-side and the sell-side for all stocks in our sample.

For the $\beta^m$ coefficients, we obtain very different results via Ridge regression (see Tables \ref{tab:ridge_results} and \ref{tab:ridge_tests}) than we do via OLS regression (see Tables \ref{tab:mlofi_ols} and \ref{tab:ols_tests}). Most notably, the values of $\beta^m$ obtained using Ridge regression have much stronger statistical significance and much smaller variance than those obtained using OLS regression. Together, these results suggest that Ridge regression does a good job at overcoming the problems caused by multicollinearity.

In Section \ref{sec:linear_MLOFI}, we refrained from performing a quantitative analysis of the values of the $\beta^m$ coefficients obtained by OLS regression, because the weak statistical significance of many of the fitted parameters may have caused us to arrive at misleading conclusions. By contrast, almost all of the fitted values of $\beta^m$ obtained by Ridge regression are strongly statistically significant (see Table \ref{tab:ridge_tests}). We therefore now turn our attention to interpreting these values.

For the small-tick stocks (i.e., AMZN, TSLA, and NFLX), the fitted values of the $\beta^m$ coefficients are approximately decreasing with increasing $m$. This suggests that for small-tick stocks, order-flow activity nearest to the bid--ask spread has the greatest impact on the contemporaneous change in mid-price, but that order-flow activity deeper into the LOB does still play a role. For the large-tick stocks (i.e., ORCL, CSCO, and MU), the fitted values of the $\beta^m$ coefficients are all quite small, but are non-zero and are strongly statistically significant. There is also no obvious decrease in the fitted values with increasing $m$ (if anything, there appears to be a slight increase). This suggests that for large-tick stocks, order-flow activity at all of the first $M=10$ price levels impacts the contemporaneous change in the mid-price. To analyze this result in more detail, we now turn our attention to assessing how well the fitted MLOFI equation \eqref{eq:mlofi_regression} captures the relationship between the net order flow at the first $M$ price levels and the contemporaneous change in mid-price, for a range of different choices of $M$.

\subsection{Assessing Goodness-of-Fit}\label{sec:gof}

To assess the goodness-of-fit of our regressions, we first follow \cite{Cont:2014price}, and consider the adjusted coefficient of determination $R^2$. We note, however, that analyzing the adjusted $R^2$ could be misleading, due to the issues with multicollinearity that we reported in Section \ref{sec:sample_correlations}. We therefore also investigate an alternative goodness-of-fit measure: the out-of-sample RMSE.

\subsubsection{Adjusted $R^2$}\label{sec:adjr2}

\begin{figure}[H] 
	\centering
	\includegraphics[width=0.7\textwidth]{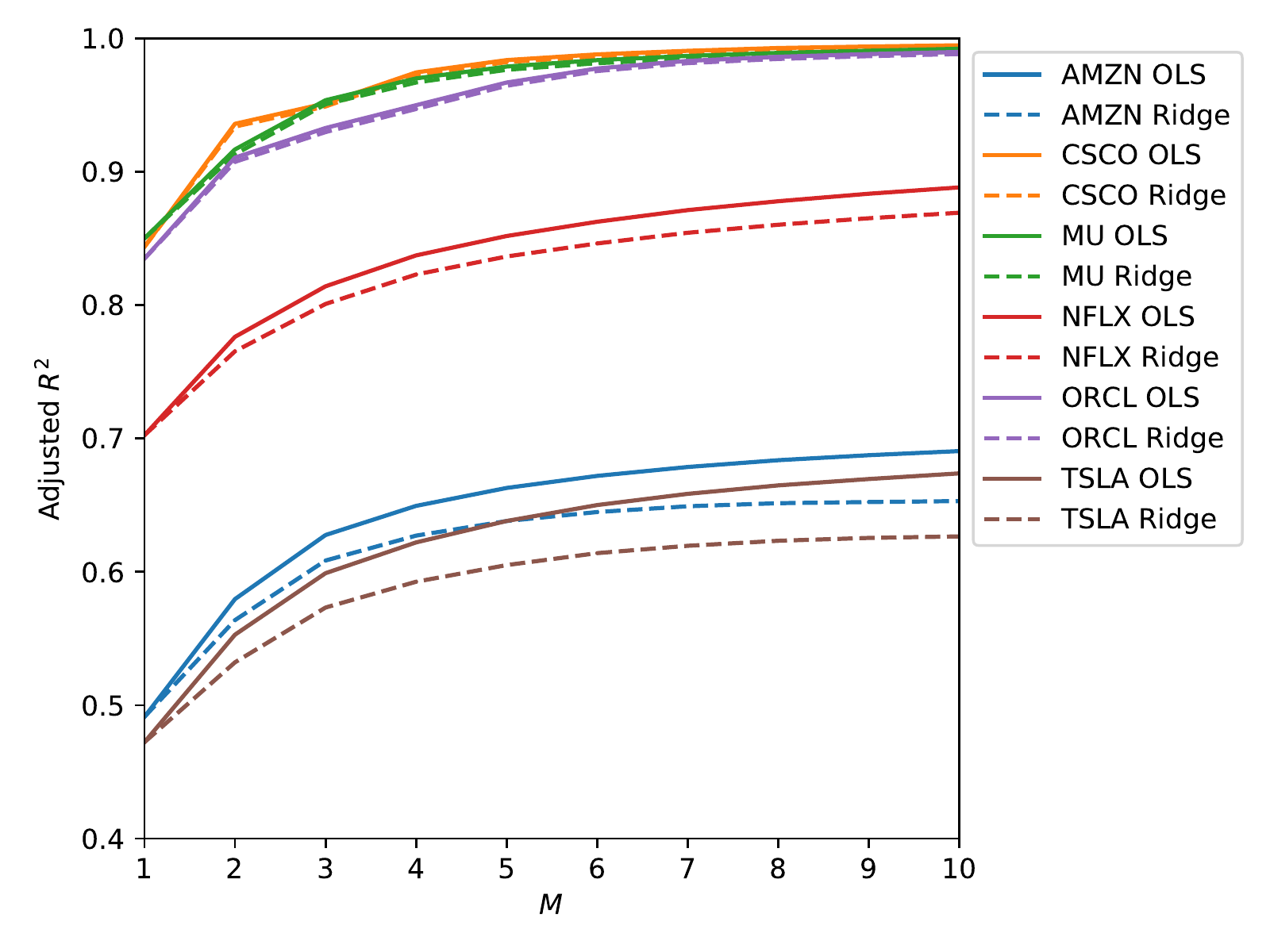}
    \caption{Mean adjusted $R^2$ statistic for various choices of $M$, for the (solid lines) OLS fits and (dashed lines) Ridge regression fits of the MLOFI equation \eqref{eq:mlofi_regression}.}
    \label{fig:adj_R2}
\end{figure}

For a given choice of $M$, the adjusted $R^2$ statistic describes the percentage of variance in the output variable (i.e., the change in mid-price in a given time window) that is explained by the first $M$ components of the MLOFI vector during the same time window, using the fitted coefficients in the MLOFI equation \eqref{eq:mlofi_regression}. Figure \ref{fig:adj_R2} shows the mean adjusted $R^2$ for each of the stocks in our sample, for $M \in \left\{1,2,\ldots,10\right\}$.

The case $M=1$ corresponds to using only net order flow at the level-1 bid- and ask-prices (i.e., the OFI equation \eqref{eq:ofi_regression}). This case is the main focus of \cite{Cont:2014price}. When $M=1$, the corresponding regression is univariate, so the output of OLS regression is the same as that of Ridge regression. The values of $R^2$ are largest for the large-tick stocks. Overall, our results are similar to those of \cite{Cont:2014price}, who reported values in the range of about $0.35$ to about $0.8$.

For all of the stocks in our sample, and when using either OLS regression or Ridge regression, the mean adjusted $R^2$ increases with $M$. Therefore, when using $R^2$ as the goodness-of-fit measure, including additional levels deeper into the LOB improves the goodness-of-fit of the MLOFI equation \eqref{eq:mlofi_regression}. The rate of increase is largest when $M$ is small, and is relatively small when $M$ is large. This suggests that for the purpose of explaining moves in the mid-price, there is indeed useful information within the net order flow at the deeper levels of an LOB, but that its impact decreases with increasing distance from the bid--ask spread.

For AMZN and TSLA (which are the smallest-tick stocks in our sample), the mean adjusted $R^2$ values for $M=10$ are about $0.65$ for OLS regression and about $0.6$ for Ridge regression; for NFLX, the mean adjusted $R^2$ values for $M=10$ are about $0.9$ for OLS regression and about $0.85$ for Ridge regression; for ORCL, CSCO, and MU (which are the largest-tick stocks in our sample), the mean adjusted $R^2$ values for $M=10$ are all very close to 1. This indicates that for large-tick stocks, the MLOFI equation with $M=10$ explains almost all of the (in-sample) variance in the change in mid-price.

For all values of $M>1$, the adjusted $R^2$ from OLS regression is greater than the corresponding adjusted $R^2$ from Ridge regression. Mathematically, this result is a direct consequence of including the penalty term in the Ridge-regression cost function \eqref{eq:ridge_regression}. In a Ridge regression, the penalty term serves to reduce the variance of the fitted parameters, but it does so at the cost of introducing a bias. By contrast, the corresponding parameter estimates produced by a standard multivariate OLS regression are unbiased, but they can exhibit a very large variance in the presence of multicollinearity, as is the case with our data (see Section \ref{sec:sample_correlations}). This could cause the OLS regression fits to perform less well out-of-sample. As we uncover in the next section, this is indeed the case.

% To assess the possible variations across different trading days, we also calculate the adjusted $R^2$ values for each trading day separately, using a selection of different values of $M$. Figure \ref{fig:amzn_daily_r} shows our results for one small-tick stock (AMZN) and one large-tick stock (ORCL); the results for all other small-tick stocks in our sample are qualitatively similar to those for AMZN and the results for all other large-tick stocks in our sample are qualitatively similar to those for ORCL.

%\begin{figure}[H] 
%	\centering
%	\includegraphics[width=0.7\textwidth]{DailyAdjustedR2.pdf}
%    \caption{Daily mean adjusted $R^2$ statistics on each trading day in 2016, for (dotted line) $M=1$, (dotted--dashed line) $M=5$, and (dashed line) $M=10$. The blue lines show the results for AMZN, which is a small-tick stock, and the purple lines show the results for ORCL, which is a large-tick stock.}
%    \label{fig:amzn_daily_r}
%\end{figure}

\subsubsection{Out-of-Sample Root Mean Squared Error}\label{sec:rmse}

Although analyzing the adjusted $R^2$ helps to provide insight into the in-sample properties of the fitted MLOFI equation \eqref{eq:mlofi_regression} for different values of $M$, this measure of goodness-of-fit also suffers from two important drawbacks in the present application. First, and most importantly, adjusted $R^2$ is an in-sample measure, in the sense that it uses the same data points to both perform the regression and to estimate the variance properties of the fitted relationship. Given the issues with multicollinearity that we described in Section \ref{sec:sample_correlations}, such an in-sample measure is likely to underestimate the true variance that would occur out-of-sample. This brings into question whether the adjusted $R^2$ is really a meaningful measure in the context of the MLOFI equation \eqref{eq:mlofi_regression}. Second, the adjusted $R^2$ measure is somewhat abstract, in the sense that it seeks to quantify the fraction of variance explained (which is a dimensionless quantity), rather than the output error, which has the dimension ``price''. This makes it difficult to interpret whether a given change in the adjusted $R^2$ is economically meaningful. To address both of these problems, we also investigate another measure of goodness-of-fit: the out-of-sample root mean squared error (RMSE).

\begin{figure}[H] 
	\centering
	\includegraphics[width=0.9\textwidth]{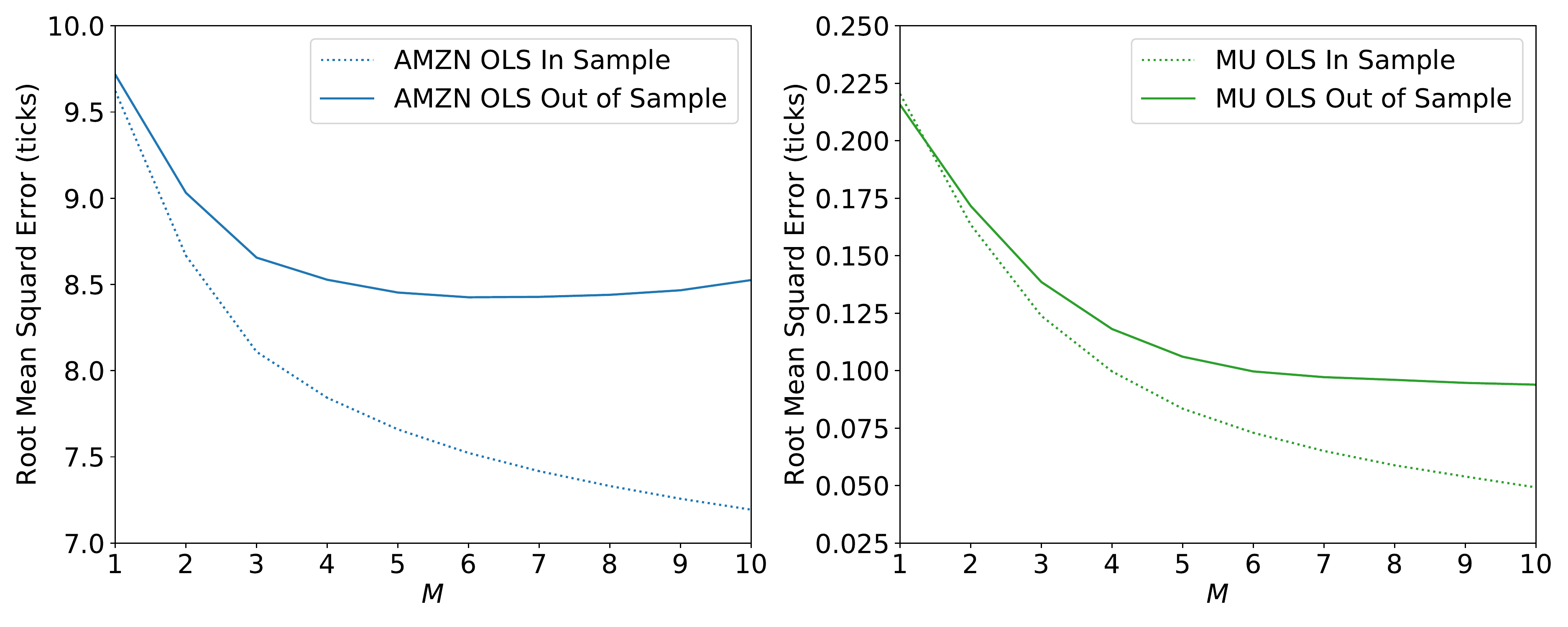}
	\includegraphics[width=0.9\textwidth]{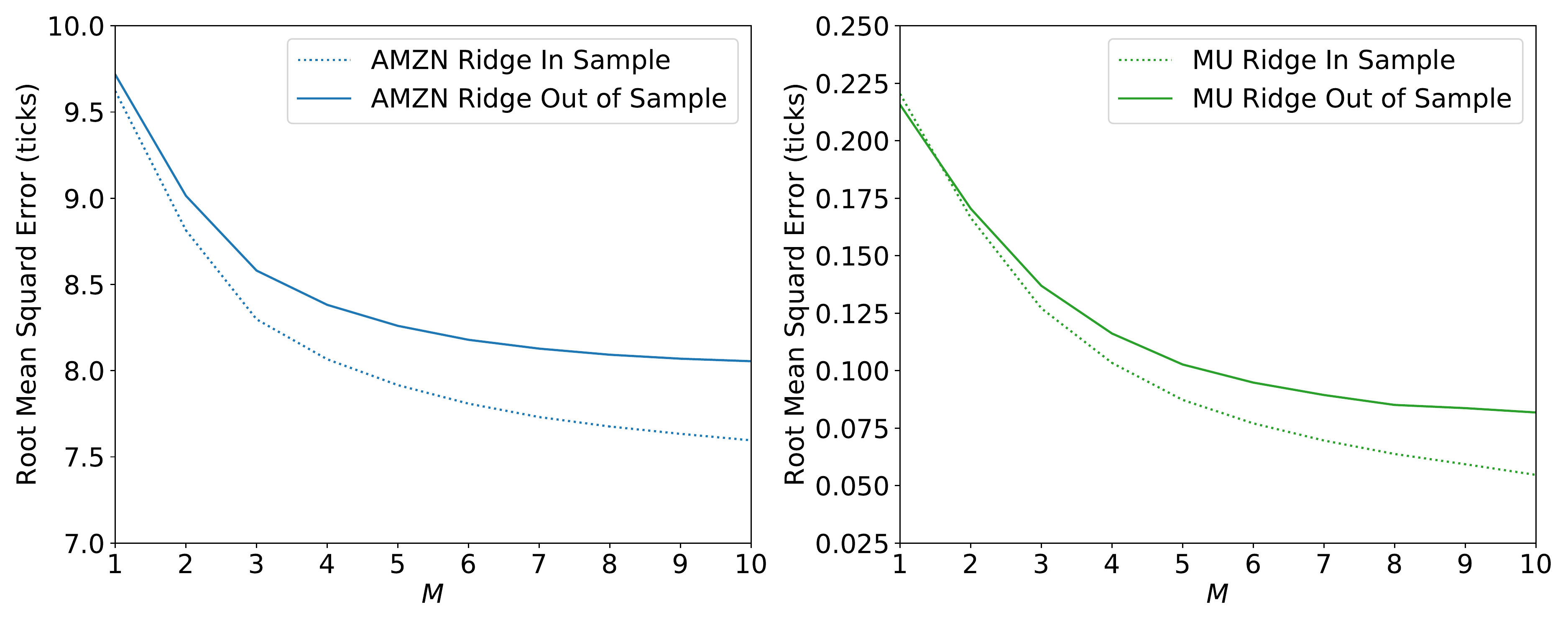}
    \caption{Mean (dotted lines) in-sample and (solid lines) out-of-sample RMSEs for (left panels) AMZN and (right panels) MU, obtained by fitting the MLOFI equation \eqref{eq:mlofi_regression} using (top row) OLS regression and (bottom row) Ridge regression. The results for TSLA are qualitatively similar to those for AMZN; the results for all the other stocks are qualitatively similar to those for MU.}
    \label{fig:is_oos}
\end{figure}

We use a methodology similar to $5$-fold cross-validation to calculate the out-of-sample RMSE. For each stock, we first split our data set into $5$ separate folds. For a given fold, we use all the data in the other 4 folds to fit the parameters of the MLOFI equation \eqref{eq:mlofi_regression} via OLS or Ridge regression. We then calculate the RMSE of the fitted MLOFI equation \eqref{eq:mlofi_regression} on these same 4 folds. We call this the \emph{in-sample RMSE}. We then use the same fitted parameters to estimate the RMSE for the other fold (which was not used in the regression fit). We call this the \emph{out-of-sample RMSE}. We repeat this process for each of the 5 folds separately, and record the mean out-of-sample RMSE across these 5 repetitions. Figure \ref{fig:is_oos} shows the mean in-sample and out-of-sample RMSEs for AMZN (which is the smallest-tick stock in our sample) and MU (which is the largest-tick stock in our sample). The results for TSLA are qualitatively similar to those for AMZN; the results for all the other stocks are qualitatively similar to those for MU.

For all stocks, and when using both OLS regression and Ridge regression, the in-sample RMSE decreases as $M$ increases. For NFLX, ORCL, CSCO and MU, when using OLS regression, the out-of-sample RMSE also decreases as $M$ increases. For AMZN and TSLA, however, the out-of-sample RMSE obtained by OLS regression first decreases, but then increases for values of $M$ larger than about 5. This is a classic hallmark of overfitting. For all the stocks in our sample, when using Ridge regression, the out-of-sample RMSE decreases as $M$ increases. This suggests that the regularization parameter in the Ridge regression cost function \eqref{eq:ridge_regression} successfully counteracts the effect of overfitting that we observe in the OLS fits for AMZN and TSLA.

\begin{figure}[H] 
	\centering
	\includegraphics[width=0.7\textwidth]{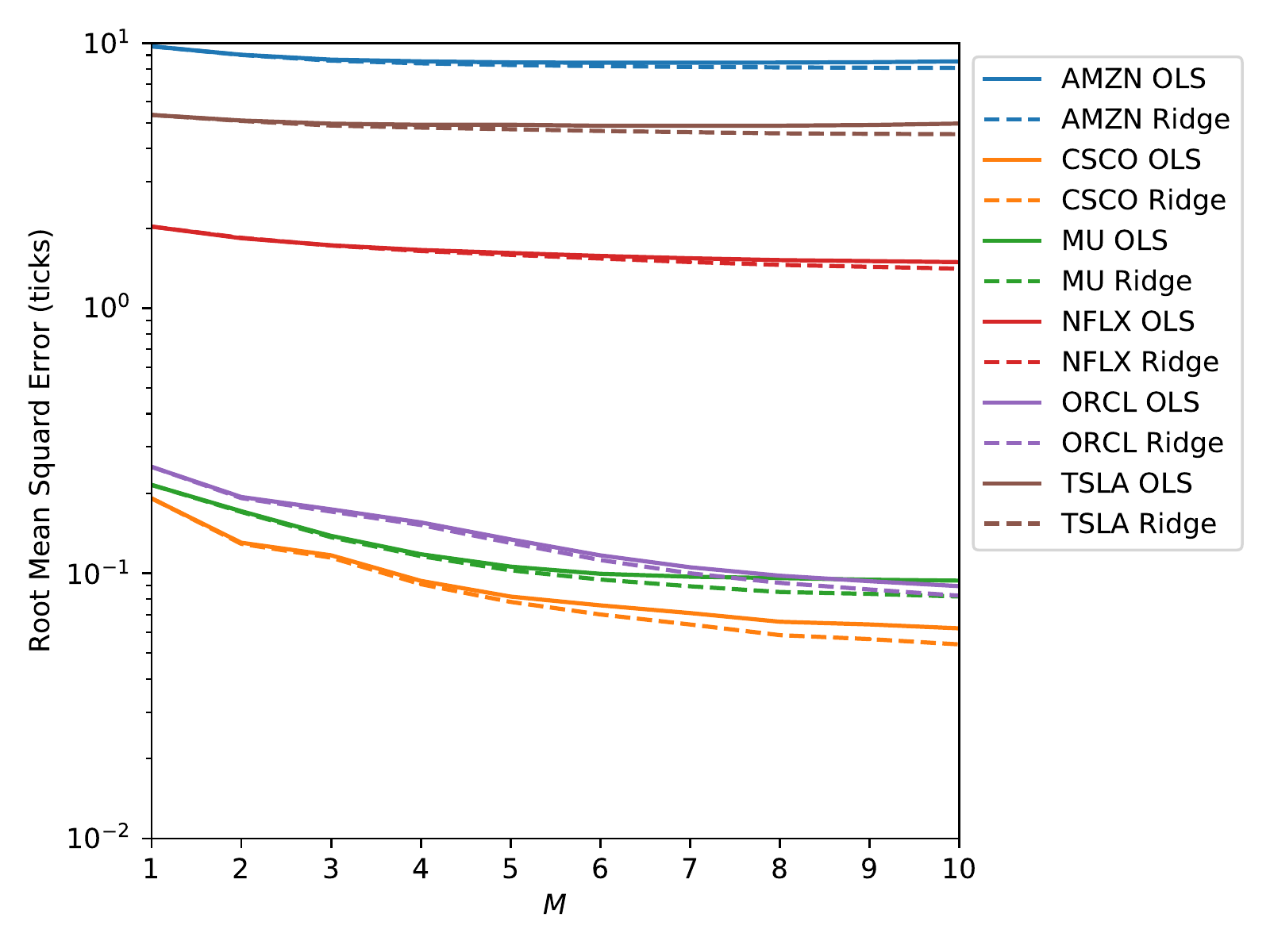}
    \caption{Mean out-of-sample RMSEs obtained by fitting the MLOFI equation \eqref{eq:mlofi_regression} using (solid lines) OLS regression and (dashed lines) Ridge regression.}
    \label{fig:RMSE}
\end{figure}

% To assess the daily variation in these results, we also calculate the out-of-sample RMSEs for each trading day separately. We perform this calculation for the OFI equation \eqref{eq:ofi_regression} and the MLOFI equation \eqref{eq:mlofi_regression} using $M=10$. Figure \ref{fig:daily_rmse} \textbf{MG to KX: Where is the data that you use to make these plots? I can't see where it is in the Dropbox, and I'd like to re-plot the figure in R.} \textcolor{red}{KX to MG: The data is in file "All other results for the rest Figures.xlsx", sheet "scores".} shows our results for one small-tick stock (AMZN) and one large-tick stock (ORCL); the results for all other small-tick stocks in our sample are qualitatively similar to those for AMZN and the results for all other large-tick stocks in our sample are qualitatively similar to those for ORCL.

%\begin{figure}[!ht] 
%	\centering
%	\includegraphics[width=\textwidth]{Figure_5_new}
%	\includegraphics[width=\textwidth]{Figure_8_new}
%    \caption{Daily RMSE obtained by using the fitted parameter values for (orange) the OFI equation \eqref{eq:ofi_regression} and (green) the MLOFI equation \eqref{eq:mlofi_regression} using $M=10$. The top panel shows the results AMZN, which is a small-tick stock, and the bottom panel shows the results for ORCL, which is a large-tick stock.}
%    \label{fig:daily_rmse}
%\end{figure}

Figure \ref{fig:RMSE} shows the mean out-of-sample RMSEs for all of the stocks in our sample, using both OLS regression and Ridge regression. As we noted in Section \ref{sec:adjr2}, when $M=1$, OLS regression and Ridge regression produce the same output. In this case, and consistently with our results when examining adjusted $R^2$, the goodness-of-fit is greatest (i.e., out-of-sample RMSE is smallest) for the large-tick stocks and is weaker for the small-tick stocks.

For all of the stocks in our sample, and when using either OLS regression or Ridge regression, the RMSE decreases with increasing $M$. Therefore, consistently with our results when examining adjusted $R^2$, we again conclude that including additional levels deeper into the LOB improves the goodness-of-fit of the MLOFI equation \eqref{eq:mlofi_regression}. The rate of improvement is again largest when $M$ is small, and is relatively small when $M$ is large. Because the unit of RMSE is ``price'', we can see that even for values of $M$ close to 10, further increase in $M$ can still reduce RMSE by a considerable fraction of a tick. In some practical scenarios, such as optimizing a high-frequency trading algorithm, an improvement of this magnitude may be economically meaningful. We return to this discussion in Section \ref{sec:discussion}.

For all stocks in our sample, and at all values of $M>1$, the out-of-sample RMSE obtained by Ridge regression is smaller than the corresponding out-of-sample RMSE obtained by OLS regression. This indicates that, when using this measure of goodness-of-fit, our Ridge regression fits outperform the corresponding OLS regression fits of the MLOFI equation \eqref{eq:mlofi_regression}. To help quantify the strength of this effect, Table \ref{table_8} shows the mean reduction in RMSE obtained by the OLS and Ridge regression fits of the MLOFI equation \eqref{eq:mlofi_regression} with $M=10$, relative to the RMSE from the fitted OFI equation \eqref{eq:ofi_regression} (i.e., relative to the fits obtained by including only the order-flow imbalance at only the level-1 bid- and ask-prices).

\begin{table} [!ht]
\begin{center}
\begin{tabular}{|P{4cm}|c|c|c|c|c|c|}
\hline
 & \textbf{AMZN} & \textbf{TSLA} & \textbf{NFLX} & \textbf{ORCL} & \textbf{CSCO} & \textbf{MU} \\ 
\hline
OFI & $9.72$ & $5.35$ & $2.03$ & $0.25$ & $0.19$ & $0.22$ \\
MLOFI, OLS & $8.53$ & $4.97$ & $1.49$ & $0.09$ & $0.06$ & $0.09$ \\
MLOFI, Ridge & $8.05$ & $4.53$ & $1.41$ & $0.08$ & $0.05$ & $0.08$ \\
\hline
Improvement of OLS MLOFI over OFI & $12\%$ & $7\%$ & $27\%$ & $64\%$ & $68\%$ & $59\%$ \\
\hline
Improvement of Ridge MLOFI over OFI & $17\%$ & $15\%$ & $31\%$ & $68\%$ & $74\%$ & $64\%$ \\
\hline
% Improvement of Ridge MLOFI over OLS MLOFI & $6\%$ & $9\%$ & $5\%$ & $11\%$ & $17\%$ & $11\%$ \\
% \hline
\end{tabular}
\end{center}
\caption{Out-of-sample RMSEs (in ticks) from stated fits of the OFI equation \eqref{eq:ofi_regression} and the MLOFI equation \eqref{eq:mlofi_regression} with $M=10$, and the corresponding improvements of these fits, measured relative to the out-of-sample RMSE achieved by OFI.}
\label{table_8}
\end{table}

In all cases, the out-of-sample RMSE obtained by using the MLOFI is smaller than that obtained by using the OFI. When using OLS regression, the improvement is about $60$--$70\%$ for large-tick stocks and about $5$--$35\%$ for small-tick stocks. When using Ridge regression, the improvement is about $65$--$75\%$ for large-tick stocks and about $15$--$30\%$ for small-tick stocks.

% The adjusted $R^2$ and the out-of-sample RMSE lead us to different quantitative conclusions about the relative goodness-of-fit achieved by OLS regression and Ridge regression. When using the adjusted $R^2$ as our goodness-of-fit measure, we conclude that OLS regression outperforms Ridge regression for all values of $M>1$. When using the out-of-sample RMSE as our goodness-of-fit measure, we draw the opposite conclusion, and instead conclude that Ridge regression outperforms OLS regression for all values of $M>1$.

% As we discussed in Section \ref{sec:rmse}, the adjusted $R^2$ is an in-sample measure of goodness-of-fit, and is therefore susceptible to underestimating the true out-of-sample variability, especially in the presence of multicollinearity. The out-of-sample RMSE does not suffer from such drawbacks. Given the strong multicollinearity that we uncovered in Section \ref{sec:sample_correlations}, we regard the results that we obtain using the out-of-sample RMSE to be a more reliable measure of goodness-of-fit. We therefore regard Ridge regression to provide better goodness-of-fit to the MLOFI equation \eqref{eq:mlofi_regression} than does OLS regression (see Table \ref{table_8}).

\section{Discussion}\label{sec:discussion}

\subsection{Comparisons to Cont \emph{et al.}}

In their study of fifty stocks from the S\&P 500, \citet{Cont:2014price} concluded that only order flow close to the best quotes has a significant impact on contemporaneous changes in the mid-price. In our study of 6 liquid stocks traded on Nasdaq, we find strong evidence to suggest that including MLOFI at price levels deep into the LOB can create a significant reduction in the out-of-sample RMSE of the fitted relationship (see Figure \ref{fig:RMSE} and Table \ref{table_8}). The reduction is especially large for large-tick stocks, which were the main focus in \citet{Cont:2014price}. This raises the interesting question of why our results differ from those reported by Cont \emph{et al.}. We propose two possible answers.

First, \citet{Cont:2014price} used OLS regression to fit their MLOFI relationship. As we discussed in Section \ref{sec:sample_correlations}, the feature variables in the MLOFI vector correspond to order-flow imbalance at neighbouring price levels within the same LOB, so it is reasonable to expect that they may exhibit multicollinearity. Indeed, this plays out empirically in our data (see Figures \ref{fig:sample_correlation} and Figure \ref{fig:eigenvalues}). This multicollinearity may cause the OLS regression fits of the MLOFI equation \eqref{eq:mlofi_regression} to be unstable, and may thereby impact the fitted values of the $\beta_m$ coefficients.

Second, \citet{Cont:2014price} used only the adjusted $R^2$ statistic to assess the goodness-of-fit of their regressions. As we discussed in Section \ref{sec:rmse}, this measure has an important drawback in the present application: It is likely to underestimate the true out-of-sample variance in the presence of muticollinearity. Although our values of $R^2$ suggests that net order flow at price levels deep into the LOB do still impact the contemporaneous change in mid-price (contrary to the findings of Cont \emph{et al.}), any conclusions based solely on this measure are susceptible to the distortions that multicollinearity may cause. If multicollinearity is especially strong in the data studied by Cont \emph{et al.}, it may yield misleading values of $R^2$.

\subsection{Comparisons to Mertens \emph{et al.}}

In a very recent working paper, \citet{Mertens:2019liquidity} extended the OFI framework from \citet{Cont:2014price} by modelling price impact as a dynamical and latent variable. Across the 5 stocks that they studied (CSCO, INTC, MSFT, CMCSA and AAPL), Mertens \emph{et al.} reported that their approach led to a reduction in out-of-sample RMSE of about $20$--$80\%$. In our own results, we achieve corresponding reductions of about $15$--$75\%$. Although we adopt very different approaches to the problem, the improvements that we achieve are within a very similar range to those achieved by Mertens \emph{et al.}. Both sets of results help to illustrate how using other input variables beyond just the OFI can help to improve the out-of-sample RMSE for the contemporaneous change in mid-price.

\subsection{Understanding Price Formation}

In an LOB, changes in the mid-price always occur as a direct consequence of changes in the bid- or ask-prices. Given that this is the case, why does including order flow at price levels deeper into the LOB yield a better goodness-of-fit for contemporaneous changes in the mid-price? We provide two possible explanations.

First, from a statistical perspective, observing a heavy in-flow of orders on the sell (respectively, buy) side of an LOB may indicate a heavy selling (respectively, buying) pressure. For example, if a trader receives private information that suggests the price is likely to rise in the medium-term, then he/she may decide to buy the asset, with the intention of holding the asset and enjoying the subsequent price increase. One way for the trader to buy the asset would be to submit a buy market order. However, by doing so, the trader reveals his/her desire to buy, and thereby risks the so-called \emph{information-leakage cost}\footnote{For a full discussion of the concept of information leakage costs, see \citet{Bouchaud:2018trades}.} that occurs due to other traders interpreting this market order arrival as a signal that reveals private information. Therefore, the trader may instead choose to submit a buy limit order, for which the associated information-leakage cost is likely to be much lower. If several different traders act in this way, and if their private information does indeed correctly anticipate a subsequent change in the mid-price, then their corresponding aggregate in-flow of limit orders will be correlated with the change in the mid-price, even if they choose to submit these limit orders at price levels deep into the LOB.

Second, from a mechanistic perspective, order flow at price levels beyond the best quotes may also influence the size of mid-price changes that occur during a given time interval. In an LOB with many limit orders at several adjacent price levels close to the best quotes, if a market order arrives and depletes the ask-queue to zero, then the ask-price can only ever increase by one tick. If, however, there is also a strong flow of cancellations at the second-best price, then it becomes more likely to observe a larger change in mid-price. As this example illustrates, all else being equal, the stronger the net out-flow of orders at prices behind the best quotes, the more likely the occurrence of a larger change in mid-price.

\subsection{Small-Tick and Large-Tick Stocks}

In our sample, half of the stocks are small-tick stocks, for which the bid--ask spread is typically several ticks wide (see Table \ref{tab:lob_means}). We observe considerable differences between our results for small-tick and large-tick stocks, including a lower statistical significance among the $\beta^m$ coefficients far from the bid--ask spread for small-tick stocks (see e.g., Table \ref{tab:ridge_results}) and a weaker goodness-of-fit for the MLOFI equation \eqref{eq:mlofi_regression}, both in terms of $R^2$ (see Figure \ref{fig:adj_R2}) and RMSE (see Figure \ref{fig:RMSE} and Table \ref{table_8}). This raises the question of why stocks with a different relative tick size should behave so differently. We provide two possible explanations. 

First, in an LOB, there are three scenarios that can cause the mid-price to change: (i) a limit order arriving inside the bid--ask spread; (ii) a sell (respectively, buy) market order consuming the whole level-1 bid-queue (respectively, ask-queue); or (iii) the last limit order in the level-1 bid- or ask-queue being cancelled. For large-tick stocks, however, the bid--ask spread is almost always at its minimum possible value of 1 tick. Whenever this is the case, it is not possible for a new limit order to arrive inside the bid--ask spread (we observed this phenomenon directly in Table \ref{tab:spread_quotes_beyond}). This prevent possibility (i) from occurring. By our definition of MLOFI, a new buy (respectively, sell) limit order arriving inside the spread will create the same MLOFI vector as a new buy (respectively, sell) limit order with the same size arriving at the level-1 bid-price (respectively, ask-price). The first such limit order arrival would create a change in mid-price, whereas the second such limit order would not. Therefore, in situations where possibility (i) can occur, the same input vectors can be mapped to different outputs. This effect may reduce the predictive power of the statistical relationship for small-tick stocks.

Second, by our definition of MLOFI, a new buy (respectively, sell) limit order arriving one tick inside the bid--ask spread will create the same MLOFI vector as a new buy (respectively, sell) limit order arriving many ticks inside the bid--ask spread. However, these events would lead to considerably different changes in the mid-price. This provides another way that the same input vectors can be mapped to different outputs when the bid--ask spread is greater than 1 tick wide. This effect may reduce the predictive power of the statistical relationship for small-tick stocks.

\subsection{Assessing Significance}

Our results in Table \ref{tab:ridge_tests} clearly demonstrate that across all of our Ridge regression fits, the vast majority of the fitted $\beta^m$ coefficients are statistically significant. However, this does not necessarily imply that their incremental impact (beyond that achieved by OFI alone) is economically meaningful. We now turn our attention to whether or not this is the case.

When studying the fitted MLOFI equation \eqref{eq:mlofi_regression}, the largest improvements in goodness-of-fit occur for small-tick stocks, and over the first few price levels (see Figure \ref{fig:RMSE}). For example, for AMZN (which is the smallest-tick stock in our sample), the incremental improvement in out-of-sample RMSE between $M=1$ and $M=3$ is more than a full tick. At the other extreme, the smallest improvements in goodness-of-fit occur for large-tick stocks, and for larger values of $M$. For MU (which is the largest-tick stock in our sample), the incremental improvement in out-of-sample RMSE between $M=9$ and $M=10$ is about $0.002$ ticks. Clearly, the magnitude of this effect is very small. However, when considering the aggregate impact of all the coefficients $\beta^2, \beta^3, \ldots, \beta^{10}$ together, the improvement in out-of-sample RMSE is much larger, and ranges from about $1.7$ ticks for AMZN to about $0.1$ ticks for MU (see Table \ref{table_8}).

Are these improvements economically meaningful? Ultimately, the answer to this question depends on context. For a trader spending $\$700$ to buy a single share of AMZN, a price difference of a couple of cents is hardly noticeable. However, for a trader operating high-frequency trading algorithms that purchase and sell huge volumes of shares many thousands of times per day, the story is quite different. Many such practitioners invest huge sums of money in the hope of improving their prediction algorithms by just a tiny fraction of a tick. Ultimately, such an improvement can make the difference between a trading strategy being profitable or unprofitable. Therefore, we argue that the improvements we report in Table \ref{table_8} are indeed economically meaningful in such cases.

%%%%%

% These results paint a different picture to the conclusions drawn by \citet{Cont:2014price}. In our own analysis, using either the adjusted $R^2$ (see Figure \ref{fig:adj_R2}) or the out-of-sample RMSE (see Figure \ref{fig:RMSE}) leads us to conclude that including net order flow from levels deeper into the LOB improves the goodness-of-fit of the MLOFI equation \eqref{eq:mlofi_regression}. In both cases, the rate of improvement is largest when $M$ is small, and is relatively small when $M$ is large. Therefore, the impact of net order flow on the contemporaneous change in mid-price decreases with increasing distance from the bid--ask spread. In \citet{Cont:2014price}, however, the authors draw the conclusion that including net order flow from levels deeper into the LOB does little to improve the goodness-of-fit. The authors make this assessment based only on their analysis of the adjusted $R^2$.

\section{Conclusions}\label{sec:conclusions}

In this paper, we performed an empirical study of the MLOFI equation \eqref{eq:mlofi_regression}, which posits a simple linear relationship between net order-flow at the first $M$ populated price levels in an LOB and the contemporaneous change in mid-price. Using recent, high-quality, high-frequency data for 6 stocks traded on Nasdaq during the period of January to December 2016, we performed both OLS regressions and Ridge regressions to fit the MLOFI equation \eqref{eq:mlofi_regression}. We used both the adjusted $R^2$ and the RMSE to assess the goodness-of-fit of our fitted relationships, and drew quantitative comparisons across our results.

By analyzing both the sample correlations in the MLOFI vector and the out-of-sample RMSE, we argued that using OLS regression to fit the MLOFI equation \eqref{eq:mlofi_regression} produces misleading results, and that Ridge regression is much better suited to this task.

When using either $R^2$ or out-of-sample RMSE, we found that the goodness-of-fit of the fitted MLOFI equation \eqref{eq:mlofi_regression} was considerably stronger for large-tick stocks than it was for small-tick stocks. For all stocks in our sample, we found that the overall goodness-of-fit increased as $M$ increased. We found that the rate of increase was largest for small $M$, and that it decreased as $M$ increased. We argued that in some cases, such as optimizing a high-frequency trading algorithm, improvements of the magnitude we observed are economically meaningful.

An obvious avenue for future research is the question of how to improve the goodness-of-fit of the MLOFI equation \eqref{eq:mlofi_regression}. As we discussed, one possible way forward might be to refine the definition of the MLOFI vector to address some of the weaknesses that we described in Section \ref{sec:discussion}. Another possible approach might be to use the MLOFI vector to model price impact as a dynamical and latent variable, using the framework recently introduced by \citet{Mertens:2019liquidity}. A third possibility might be to include other input variables into the regression.

Another possible avenue for future research is to develop a deeper understanding of the differences between small-tick and large-tick stocks. In some sense, it may seem inevitable that stocks with different tick sizes should behave differently. However, many empirical and theoretical studies of LOBs have suggested that by performing appropriate re-scalings, many seemingly idiosyncratic properties of specific LOBs may actually be universal (see, e.g., \citet{Bouchaud:2002statistical,Bouchaud:2018trades,patzelt2018universal,potters2003more}). It would be interesting to understand whether implementing rescaling in either the inputs (i.e., the MLOFI vector) or the output (i.e., the change in mid-price) of the MLOFI equation \eqref{eq:mlofi_regression} could uncover universal behaviours across vastly different LOBs.

\appendix

\section*{Appendix A: Intra-Day Seasonalities}\label{app:intraday}

As we discuss in Section \ref{sec:intraday}, both \citet{Cont:2014price} and \citet{Mertens:2019liquidity} reported intra-day seasonalities in the OFI relationships that they studied. To help assess the extent to which intra-day seasonalities may have impacted our own results, we also studied how the values of the fitted $\beta^m$ parameters vary throughout the trading day.

\begin{figure}[H] 
	\centering
	\includegraphics[width=0.9\textwidth]{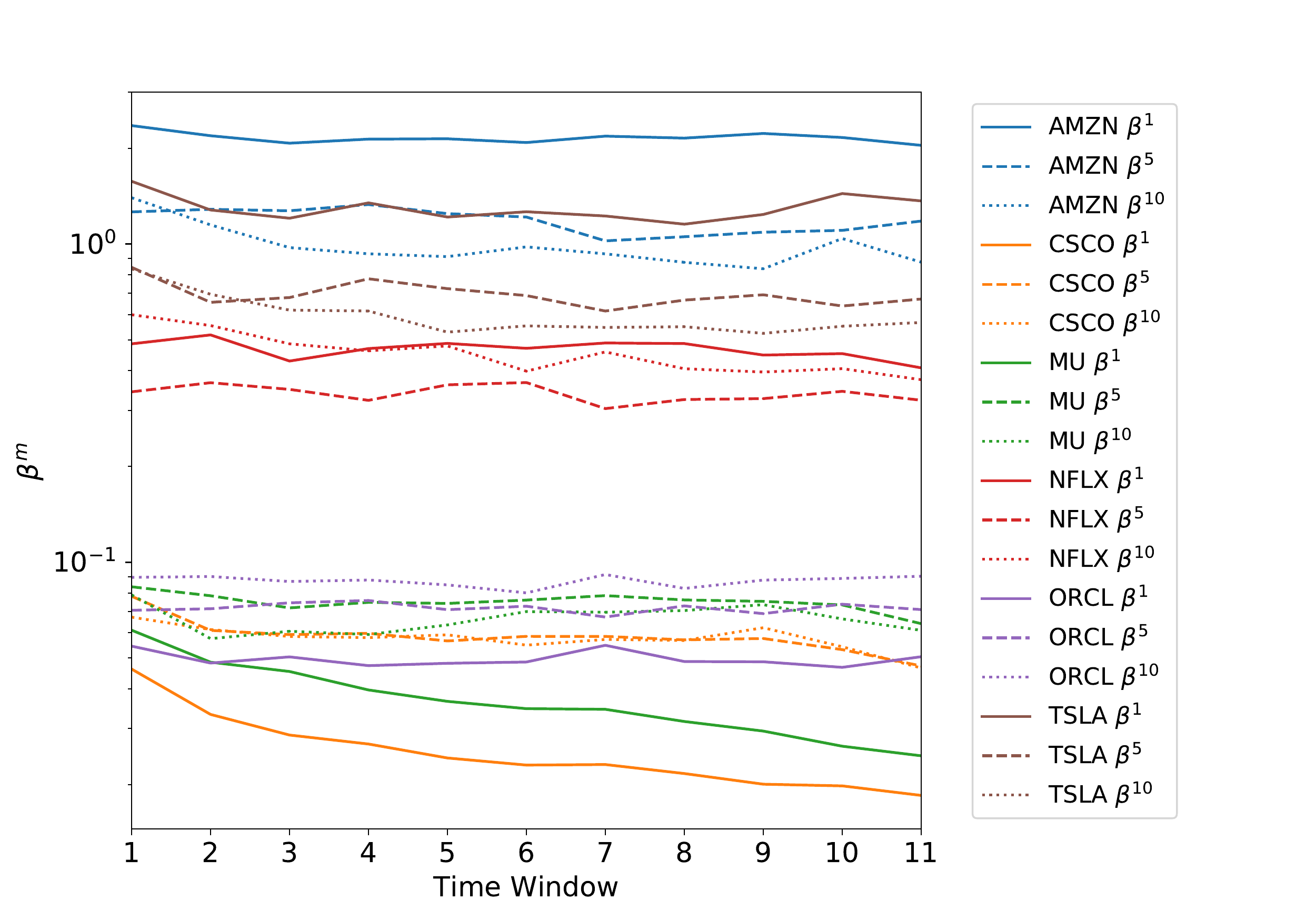}
    \caption{Ridge regression estimates of the mean (solid lines) $\beta^1$, (dashed lines) $\beta^5$, and (dotted lines) $\beta^{10}$ coefficients in the MLOFI equation \eqref{eq:mlofi_regression}, when taking the average across the corresponding fitted values using only the data from the given 30-minute time window. The $1^{\text{st}}$ time window runs from $10$:$00$--$10$:$30$, the $2^{\text{nd}}$ time window runs from $10$:$30$--$11$:$00$, and so on. The results when fitting the parameters by OLS regression are qualitatively similar.}
    \label{fig:intraday}
\end{figure}

To perform this analysis, we used the same sample construction methodology as for our main calculations (see Section \ref{sec:sample_construction}), except that we calculated the mean fitted value of the $\beta^m$ parameters for each of the $I=11$ different windows separately. For example, when examining the $\beta^1$ parameter for AMZN during the first time window (which runs $10$:$00$--$10$:$30$), we calculated the mean among the 252 fitted values of the $\beta^1$ parameters when using only the data from $10$:$00$--$10$:$30$ each day. We performed this analysis for both OLS regression and Ridge regression. Figure \ref{fig:intraday} shows the mean values of the $\beta^1$, $\beta^5$, and $\beta^{10}$ parameters fitted using Ridge regression during each of the $I=11$ different windows. The behaviour of the other $\beta^m$ parameters is qualitatively similar to that for $\beta^5$, and $\beta^{10}$. The results when fitting the $\beta^m$ parameters using OLS regression are also qualitatively similar.

For CSCO and MU, which are the two largest-tick stocks in our sample, we observe a clear intra-day decrease in the mean fitted values of the $\beta^1$ parameters. This suggests that for a given net order flow at the level-1 bid- and ask-prices, the contemporaneous change in mid-price is typically smaller during time windows that occur later in the trading day. This result is consistent with both \citet{Cont:2014price} and \citet{Mertens:2019liquidity}, who also focussed primarily on large-tick stocks. For all of the other stocks in our sample, however, we find that the magnitude of this effect is (at best) very small, and is difficult to separate from the natural variability that occurs due to random fluctuations. For all of the stocks in our sample, we observe no systematic intra-day decrease in the mean fitted values of any of the other $\beta^m$ parameters (i.e., $\beta^2$, $\beta^3$, \ldots, $\beta^{10}$). Therefore, given that this effect only occurs for the $\beta^1$ parameter for 2 of our stocks, and given also that intra-day seasonalities are not the core focus of our work, we choose to present all of our empirical calculations in the main body of the paper when using all $I=11$ time windows together.

\section*{Acknowledgements}We thank an anonymous reviewer for many helpful comments and suggestions. We thank EPSRC and the James S. McDonnell Foundation for supporting this research.

\bibliographystyle{plainnat}
\bibliography{ref}

\end{document}